\documentclass[12pt]{article}
\pdfoutput=1 
\usepackage{putex}
\usepackage{feyn}
\usepackage[vcentermath]{youngtab}
\usepackage{amssymb,amsmath,amsthm,mathrsfs}
\usepackage{graphicx}
\usepackage{epstopdf}
\usepackage{enumerate}
\usepackage{cite}
\usepackage{tensor}
\usepackage{slashed}
\usepackage{feynmf}

\usepackage{hyperref} 
 
\numberwithin{equation}{section}

\newcommand{\es}[2] {\begin{equation} \label{#1} \begin{split} #2 \end{split} \end{equation}}
\newcommand{\eq}[2] {\begin{equation} \label{#1} #2 \end{equation}}

\newcommand{\Z}{\mathbb{Z}}
\newcommand{\N}{\mathbb{N}}
\newcommand{\R}{\mathbb{R}}
\newcommand{\CC}{\mathbb{C}}

\newcommand{\cO}{\mathcal{O}}

\newcommand{\cF}{\mathcal{F}}

\def\half{{1\over  2}}

\newcommand{\be}{\begin{equation}}
\newcommand{\ee}{\end{equation}}

\newcommand{\bb}{\mathbb}
\newcommand{\cl}{\mathcal}
\newcommand{\p}{\partial}

\def\e{{\rm e}}
\def\cO{{\cal O}}

\def\vac{{\rm vac}}

\def\vs{\vskip .1 in}
\def\subsec{\subsection}
\def\N{{\cal N}}

\def\C{{\cal C}}
\def\cF{{\cal F}}

\def\L{\Lambda}

\def\eqr{\eqref}

\def\b{\beta}
\def\l{\lambda}
\def\t{\tau}

\def\bul{$\bullet$\quad}
\def\s{\sigma}
\def\Eh{\hat E}
\def\Oc{{\cal O}}
\def\vs{\vskip .1 in}
\def\S{\Sigma}
\def\Chh{C_{h_1,h_2}(x)}
\newcommand{\scr}{\mathscr}
\def\R{{\scr R}}

\newcommand{\bea}{\begin{eqnarray}}
\newcommand{\eea}{\end{eqnarray}}

\def\Tr{\mathop{\rm Tr}}

\newcommand\Om{\Omega}

\def\rar{\rightarrow}

\newcommand{\ev}[1]{\langle{#1}\rangle}

\begin{document}

\preprint{BRX-TH-6292}

\institution{BU}{Martin Fisher School of Physics, Brandeis University, Waltham, MA 02454, USA}
\institution{MG}{Department of Physics, McGill University, Montr\'eal, QC H3A 2T8, Canada}
\institution{PU}{Department of Physics, Princeton University, Princeton, NJ 08544, USA}

\title{R\'enyi Entropies, the Analytic Bootstrap, and 3D Quantum Gravity at Higher Genus}

\authors{Matthew Headrick\worksat{\BU}, Alexander Maloney\worksat{\MG}, Eric Perlmutter\worksat{\PU}, Ida G. Zadeh\worksat{\BU}
}

\abstract{
We compute the contribution of the vacuum Virasoro representation to the genus-two partition function of an arbitrary CFT with central charge $c>1$.  This is the perturbative pure gravity partition function in three dimensions. We employ a sewing construction, in which the partition function is expressed as a sum of sphere four-point functions of Virasoro vacuum descendants. For this purpose, we develop techniques to efficiently compute correlation functions of holomorphic operators, which by crossing symmetry are determined exactly by a finite number of OPE coefficients; this is an analytic implementation of the conformal bootstrap. Expanding the results in $1/c$, corresponding to the semiclassical bulk gravity expansion, we find that---unlike at genus one---the result does not truncate at finite loop order. Our results also allow us to extend earlier work on multiple-interval R\'enyi entropies and on the partition function in the separating degeneration limit. 
}
 
\date{}
\maketitle

\tableofcontents

\section{Introduction}

Three-dimensional gravity has proven to be a fruitful testing ground for our ideas about holography. 
Perhaps the most interesting question is whether pure general relativity---a theory with only metric degrees of freedom---with a negative cosmological constant exists as a quantum theory in its own right.  If this were the case, then one should be able to find its holographic dual for a given value of $G_N/R_{\rm AdS}$, the Newton constant in AdS units.  This appears to be an extremely difficult problem (see e.g. \cite{Witten:2007kt, Maloney:2007ud, Castro:2011zq}).  However, general relativity exists as a sub-sector of any theory of gravity in three dimensions.  From the boundary point of view, it captures the dynamics of the Virasoro sector of any two-dimensional CFT with central charge $c=3R_{\rm AdS}/2G_N$. This semi-microscopic interpretation is unavailable in higher-dimensional AdS/CFT, where the stress tensor does not generate a closed symmetry algebra. 

This perspective lends a universality to AdS$_3$/CFT$_2$ that underlies, for example, the matching of the asymptotic symmetry group of anti-de Sitter space to the Virasoro algebra \cite{Brown:1986nw} and the matching of BTZ black hole entropy to the Cardy growth of states \cite{Strominger:1997eq}.  These are features of any theory of three-dimensional gravity, and of any dual CFT.  Recent work has revealed an even richer set of properties of two-dimensional CFTs that admit a large-$c$ limit and are dual to weakly coupled bulk theories of gravity. These relate to aspects of such theories' spectra and thermodynamics \cite{Hartman:2014oaa, Keller:2014xba, Belin:2014fna, Haehl:2014yla, Benjamin:2015hsa}, entanglement \cite{rt, Headrick:2010zt, Hartman:2013mia, Faulkner:2013yia, Barrella:2013wja, Chen:2013kpa, Castro:2014tta, Asplund:2014coa, Caputa:2014eta}, Virasoro blocks \cite{Fitzpatrick:2014vua, deBoer:2014sna, Hijano:2015rla, Fitzpatrick:2015zha, Perlmutter:2015iya}, modular geometry \cite{Jackson:2014nla}, and chaotic response to perturbations \cite{Roberts:2014ifa}, among others.

Despite this fascinating progress, much remains to be understood about basic consequences of Virasoro symmetry. To this end, in this paper we will focus on the computation of the partition function of three-dimensional gravity in a universe whose boundary is a Riemann surface $\S_g$ of genus $g$.  Schematically, this should be given by a bulk path integral over geometries ${\cal M}_g$ which asymptote to $\S_g$:
\es{in2}{Z_{\rm grav}(\Om_g) = \int_{\p {\cal M}_g = \S_g} {\cal D}g~ e^{-S[g]}~.}
This partition function is a function of the conformal structure moduli of the Riemann surface $\S_g$, denoted $\Om_g$. These partition functions contain vital information about the theory: for instance, one can recover the correlation functions of a given CFT from its higher-genus partition functions by pinching handles \cite{Friedan:1986ua}. Thus, by tuning the moduli $\Om_g$, one could in principle recover the correlation functions of the boundary CFT.

Equation \eqr{in2} is in general an extremely difficult object to compute.  Moreover, it is not ``universal'' in the sense described earlier. 
In particular, in \eqr{in2} we have written the bulk path integral only over metric degrees of freedom; in more complicated theories of gravity more degrees of freedom should be included.
The partition function $Z_{\rm grav}(\Om_g)$ written above is that of the CFT dual to pure gravity at a given value of Newton's constant, if it exists. 
In this paper we will not be interested in the full partition function \eqr{in2}, but rather in an object which is both easier to compute and universal: we will study the contribution to $Z_{\rm grav}(\Om_g)$ from a single saddle-point geometry ${\cal M}_g$, including perturbative quantum corrections.  This restricted partition function maps to the contribution of the Virasoro sector to the CFT partition function on $\S_g$. 

This is easiest to see at genus one. In the semiclassical regime $G_N \ll R_{\rm AdS}$, the path integral \eqr{in2} can be recast as a sum over saddle points of the Einstein action with solid torus topology, along with perturbative corrections. The simplest such saddle point is thermal AdS$_3$, the Euclidean geometry found by taking empty AdS$_3$ and periodically identifying in Euclidean time, which contributes to the partition function as
\es{in3}{Z_{\rm TAdS}(\t,\overline{\t}) = |q|^{-c/12} \prod_{n=2}^{\infty}{1\over |1-q^n|^2}~,\quad  q:=e^{2\pi i \t}~.}
In this expression, we have included not only the classical action of thermal AdS (the factor of $|q|^{-c/12}$) but also all of the perturbative quantum corrections which come from loops of gravitons in thermal AdS.
With certain reasonable assumptions, all other saddle points are simply $SL(2,\mathbb{Z})$ modular transformations of thermal AdS, and the sum over geometries is a  sum over $SL(2,\mathbb{Z})$ transformations of \eqr{in3}. A direct calculation of $Z_{\rm TAdS}$ does not yield a result consistent with an interpretation as a trace of the Hilbert space of a CFT \cite{Maloney:2007ud, Keller:2014xba}.\footnote{However, in the quantum regime $G_N \sim R_{\rm AdS}$, it was argued in \cite{Castro:2011zq} that at specific minimal-model values of $c$, not only can the sum be performed, but it agrees with the minimal-model partition functions.} Nevertheless, \eqr{in3} does have a natural interpretation as the Virasoro vacuum character of {\it any} CFT with central charge $c>1$ and an $SL(2,\mathbb{R}) \times SL(2,\mathbb{R})$-invariant ground state.  We note that this object is not modular invariant, which reflects the fact that in \eqr{in3} we have focused on only one saddle out of the $SL(2,\Z)$ family.  In the language of Riemann surfaces, \eqr{in3} is a function not of the conformal structure of the boundary torus, but rather of the Teichm\"uller parameter $\tau$.

In the present paper, our goal is to compute the analog of \eqr{in3} at higher genus. Any theory of AdS$_3$ gravity contains solutions which are handlebodies---solid genus-$g$ geometries---which have the Riemann surface $\S_g$ as their conformal boundaries.  These solutions are quotients of Euclidean AdS$_3$, much like thermal AdS.  The contribution of a given handlebody to the path integral---including graviton loop corrections---has a universal CFT interpretation for any value of $c$, as the contribution of the states in the vacuum representation to the CFT partition function on $\S_g$. We call this quantity $Z_{\rm vac}$.\footnote{This was called $Z_{fake}$ in \cite{Yin:2007gv}, and the correspondence with the bulk saddle point partition function was written as $Z_{fake} = Z_{saddle}$.}   $Z_{\rm vac}$ is a function not just of the conformal structure of $\S_g$, but rather of the Teichm\"uller parameters that parametrize the universal cover of the moduli space. In other words, to compute $Z_{\rm vac}$ we must specify a marking of the Riemann surface $\S_g$ that fixes a choice of contractible and non-contractible cycles of the handlebody (A- and B-cycles, respectively).  Thus $Z_{\rm vac}$ is not invariant under the modular group; a modular-invariant partition function could be obtained only, for example, by summing over bulk saddles that describe the different ways a boundary $\S_g$ can be ``filled in" by bulk geometries. One of our goals in this paper will be to give a direct CFT computation of 
$Z_{\rm vac}$, which can then be interpreted gravitationally.

There has been a recent resurgence of interest in higher-genus partition functions of two-dimensional CFTs. This interest is partly motivated by the study of entanglement entropies (EEs).  The computation of EEs via the replica trick involves evaluating entanglement R\'enyi entropies (EREs), which in turn are equal to certain higher-genus partition functions. A particularly interesting line of research uses calculations of EREs to test the Ryu-Takayanagi (RT) classical EE formula and understand the quantum corrections to it \cite{rt, Hartman:2013mia, Faulkner:2013yia, Barrella:2013wja, Faulkner:2013ana}.

The partition functions relevant for EREs have been computed in holographic CFTs in two ways: in gravity, by explicitly finding the relevant saddles and evaluating their classical actions and one-loop determinants, and in field theory, by computing twist-operator correlators in certain cyclic orbifold CFTs and then expanding the results in powers of $1/c$. In every case where the computation has been carried out on both sides, agreement has been found. This is a check of our basic understanding of AdS$_3$/CFT$_2$ duality. Further, in many cases the computation using one technique gives results that are not practically computable using the other, thereby giving new information about both three-dimensional gravity and large-$c$ CFTs. For example, by expanding the results of the twist-operator computation to higher orders in $1/c$, one determines higher-loop quantum corrections on the gravity side that would be exceedingly difficult to obtain by direct computation.  These results hint at a surprising novel structure, which we describe below.  

\subsec{Summary of results}

\begin{figure}
\centering
\includegraphics[width=0.98\textwidth]{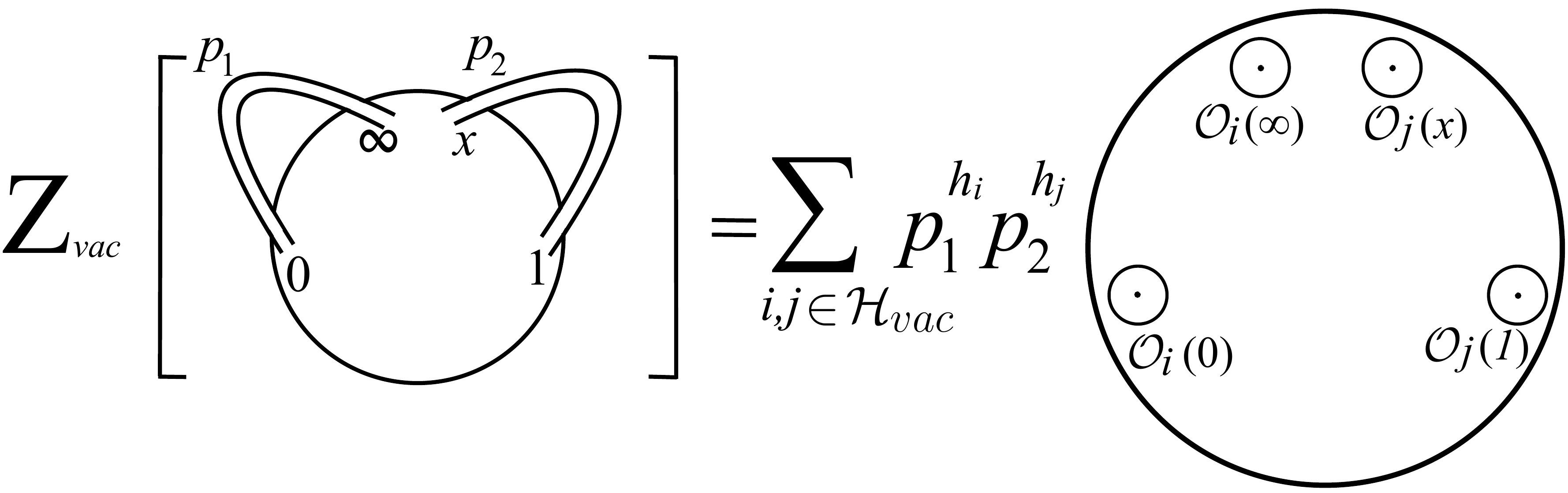}
\caption{\label{fig-sewing-vac}A depiction of the sewing construction as applied to $Z_{\rm vac}$, the contribution of the Virasoro vacuum representation to a genus-two CFT partition function. The coordinates $p_1$ and $p_2$ represent the widths of the two handles in a Schottky uniformization of the Riemann surface. The handles are replaced by a sum over pairwise operator insertions, where we include all Virasoro descendants of the identity, $\Oc \in {\cal H}_{\rm vac}$. This recasts $Z_{\rm vac}$ as a sum of sphere four-point functions, weighted by powers of $p_1$ and $p_2$. The operators $\cO_i$ and $\cO_j$  have holomorphic conformal weights $h_i$ and $h_j$, respectively. A detailed description of the sewing construction is presented in section \ref{iv} (see equations \ref{Z2} and \ref{Ch1h2}).}
\end{figure}

In this paper we will directly compute $Z_{\rm vac}$ at genus two, for arbitrary values of $c>1$, using CFT techniques. We will use a sewing construction, represented schematically in figure \ref{fig-sewing-vac}.  We start with a Riemann surface $\S$ that has been constructed by Schottky uniformization and replace the handles of $\S$ with a sum over states propagating along these handles. The result is a weighted sum over four-point functions of local  operators on the sphere.  If we were to include all possible operators in this sum, we would obtain the full, modular-invariant partition function, as a function of the pinching parameters $p_1$ and $p_2$, that describe the widths of the handles, and a third modulus $x$, which is the cross-ratio of the four-point functions on the sphere.  The universal contribution $Z_{\rm vac}$ is computed by summing only over operators in the Virasoro vacuum block. The four-point functions of these operators are determined completely by conformal Ward identities.  Thus $Z_{\rm vac}$ is in principle completely determined in terms of the central charge.  We will compute the answer perturbatively in $p_1$ and $p_2$ but exactly in $x$. We will mostly assume that the CFT has no extended symmetry algebra beyond two copies of Virasoro, although as we will see in section \ref{v-i-i} it is straightforward to extend our results to higher-spin theories.

Conformal bootstrap methods play an important role in our computation of $Z_{\rm vac}$, since our computation requires us to sum over all four-point functions of Virasoro descendants on the sphere.  These correlation functions---and indeed the correlation function of any family of chiral operators---can be efficiently computed using a holomorphic version of the conformal bootstrap.  The essential idea is that these correlation functions can be regarded as a meromorphic functions on ${\cal M}_{0,4}$, the moduli space of four marked points on a sphere, with poles only when the operators coincide.  A meromorphic function on a compact space is determined entirely by its polar behaviour.  For chiral operators of finite conformal dimension, this polar part is determined by a finite number of three-point function coefficients.  The result is an exact expression for the correlation function in terms of a {\it finite} number of three-point function coefficients.  This should be contrasted with the usual approach, where a four-point function involves a sum over an infinite number of intermediate states, so is written in terms of an infinite sum of OPE coefficients.  Similar ideas have been advanced in \cite{Bouwknegt:1988sv, Bowcock:1990ku, Keller:2013qqa}. When the chiral operators are Virasoro descendants of the identity, we show using free bosons that all connected $n$-point functions have polynomial dependence on $c$. This implies that, when expressed in terms of $c$, bulk scattering of graviton states in AdS$_3$ is purely classical, in analogy to the one-loop exactness of the torus partition function.

Our result for $Z_{\rm vac}$ will hold for a general Riemann surfaces $\S$, but for certain values of the moduli---those corresponding to the so-called replica surface---our results can be used to compute genus-two EREs.  We mainly consider the case of two disjoint intervals in the vacuum of a CFT; the replica manifold has genus two when $n=3$, and is denoted $\R_{2,3}$. Our results extend previous results in \cite{Chen:2013kpa, Chen:2013dxa}, which were obtained from the twist-field four-point function. Those works employ a short-interval expansion in which the conformally invariant cross-ratio, which we call $y$, is taken to be small. The sewing technique is well-suited to computation to higher orders in $y$; \cite{Chen:2013dxa} worked through $O(y^8)$, and we extend this to $O(y^{12})$. 
In fact, the authors of \cite{Chen:2013dxa} found a quite remarkable result: their $O(y^8)$ term in $\log Z_{\rm vac}$ exhibits a two-loop truncation in the expansion in $1/c$ at large $c$.

To understand why that result is interesting, let us consider the bulk AdS$_3$ interpretation of our results.  Our computation of $Z_{\rm vac}$ is not limited to large $c$; it is a truly quantum result for the saddle-point partition function for a genus-two handlebody of three-dimensional pure gravity, applicable even when $G_N/R_{\rm AdS}$ is of order one. Expanding our result at large $c$ is equivalent to making the semiclassical approximation in the bulk.  More precisely, the expansion of the ``vacuum free energy,''%
\be\label{in5}
F_{\rm vac} := -\log Z_{\rm vac}\,,
\ee
 at small Newton constant (large $c$) is the loop expansion of three-dimensional AdS gravity:
\es{in6}{F_{\rm vac} = \sum_{\ell=0}^{\infty}c^{1-\ell} F_{\rm vac;\,\ell}}
where $F_{\rm vac;\,\ell}$ denotes an $\ell$-loop contribution. In the bulk, no computations have been done beyond one-loop order. At one loop, there is a closed-form expression for the graviton handlebody determinant \cite{Yin:2007gv, Giombi:2008vd}. Our result for $F_{\rm vac;1}$ is a computation of this determinant in a new regime of moduli space, not described by previous computations \cite{Yin:2007gv, Barrella:2013wja}.
 
What about at higher loops? At genus one, the expansion \eqr{in3} truncates at one-loop order: higher-loop contributions only renormalize the value of $c$ \cite{Maloney:2007ud}. It is natural to ask whether the higher-genus partition function obeys an analogous truncation. 
Indeed, the results of  \cite{Chen:2013dxa}  imply that
that $F_{\ell>2}(\R_{2,3})$ vanishes through $O(y^8)$, perhaps suggesting that the partition function at genus two truncates at two loops.  One motivation for this paper was to investigate whether this truncation really occurs for the full partition function $Z_{\rm vac}$.\footnote{
On general grounds, such a truncation might seem to conflict with the pole structure of CFT correlation functions, regarded as analytic functions of $c$. Let us make the argument at genus two for concreteness. In the sewing construction, a genus-two partition function is written as a sum over four-point functions. The statement of truncation becomes the statement that at each order in the sewing expansion, the total contribution from all four-point functions truncates at order 1/c. As argued by Zamolodchikov \cite{zamo}, the conformal block decomposition of a given four-point function contains poles at minimal model values of $c$ where the exchanged operators become null. Unless these poles cancel against the poles in the other four-point functions contributing at a given order in the sewing expansion, the partition function will not truncate in a $1/c$ expansion. This sort of cancellation at every order in the sewing expansion seems highly unlikely.  Indeed, our computations bear out this conclusion.}  
Our conclusion is that the truncation does not occur, and that the cancellation observed in \cite{Chen:2013dxa} is an artifact of the small-$y$ expansion. 
Indeed, we will show that  on the replica manifold $\R_{2,3}$ there are nonzero contributions to the free energy at all orders in the $1/c$ expansion. These first appear at $O(y^{12})$ in the short-interval expansion, explaining why these corrections were not found in \cite{Chen:2013dxa}.  

More generally, we will show that the genus-two partition function $Z_{\rm vac}$ of pure three-dimensional gravity does not truncate at any order in $1/c$. The same is true of pure higher spin gravity. Explicit contributions to the all-loop terms $F_{\rm vac;\;\ell}$ are given in section\, \ref{v-i}. To our knowledge, these are the first all-loop results beyond genus one for a Riemann surface with three independent moduli. We show that in the regime of small $p_1$ and $p_2$, the only point in the moduli space at which the loop expansion \eqr{in6} truncates is the separating degeneration point, at which $\S$ degenerates into the union of two tori.

This paper is organized as follows. In section \ref{ii} we recall the relationship between R\'enyi entropies and higher-genus partition functions, and review the sewing construction of higher genus partition functions as a weighted sum over sphere-correlation functions.  In section \ref{iii} we describe techniques to compute these correlation functions, including an analytic version of the conformal bootstrap.  In section \ref{iv} we apply these techniques to compute $Z_{\rm vac}$, the contribution to the partition function from the Virasoro descendants, at genus two.  We discuss the large central charge limit of this result in section \ref{v}, which allows us to understand the nature of quantum corrections to the higher-genus partition function of three-dimensional gravity, as well as applications to R\'enyi entropies, before concluding in section \ref{vi}. Appendices contain details relevant to the sewing construction.

\section{Review}\label{ii}

In this section we will review some relevant background material, and explain the methodology and philosophy behind our computations. In subsection \ref{Renyireview}, we briefly review previous work on R\'enyi entropies in the vacuum of 2D CFTs. Using these R\'enyi entropies as a guide, we explain how pure 3D quantum gravity naturally computes the universal contribution of the Virasoro identity block to CFT partition functions on generic Riemann surfaces. Then, in subsection \ref{sewing}, we will explain the sewing construction, which we will apply in section \ref{iv} to the computation of higher-genus partition functions.

\subsection{R\'enyi entropies and higher-genus partition functions}\label{Renyireview}
Two-dimensional CFTs provide perhaps the simplest arena in which to investigate entanglement entropies (EEs) in field theories. In this subsection, we will briefly review some calculations of these quantities, with particular attention to their dependence on the central charge $c$ of the theory.

\subsubsection{General CFTs}

The simplest quantity one can consider in this context is the vacuum EE of a single interval $[u,v]$ on the line. The corresponding R\'enyi entropy is given in terms of the partition function on the surface\footnote{In general, we will denote the plane branched $n$ times over a set of $N$ intervals by $\R_{N,n}$; this surface has genus $(N-1)(n-1)$.} $\R_{1,n}$, which is the plane branched $n$ times over the interval \cite{Holzhey:1994we}:\footnote{The partition function on a genus-zero surface is defined, for a given theory, up to a multiplicative constant independent of the metric. We are choosing that constant so that $Z(\CC)=1$; otherwise the argument of the logarithm in \eqref{SZrelation} would be $Z(\R_{1,n})/Z(\CC)^n$.}
\begin{equation}\label{SZrelation}
S^{(n)}([u,v]) = -\frac1{n-1}\ln Z(\R_{1,n})\,.
\end{equation}
This partition function is in turn related to the two-point function on the plane of twist operators in the orbifold theory $\C^n/\Z_n$ (where $\C$ is the original CFT) \cite{cardy0}:
\begin{equation}
Z(\R_{1,n}) = \ev{\sigma(u)\tilde\sigma(v)}_{\C^n/\Z_n}\,.
\end{equation}
It will be convenient to work in terms of the free energy, which we define on any surface $X$ as
\begin{equation}
F(X):=-\ln Z(X)\,.
\end{equation}
The free energy $F(\R_{1,n})$ is proportional to $c$ and otherwise independent of the theory: the surface $\R_{1,n}$ has genus zero, so the free energy is given entirely by a Liouville action multiplied by $c$; alternatively, in the twist-operator language, the two-point function depends only on their dimension, which is proportional to $c$. The result is \cite{Holzhey:1994we,cardy0}
\begin{equation}
S^{(n)}([u,v]) = \frac c6\left(1+\frac1n\right)\ln\left(\frac{v-u}\epsilon\right),
\end{equation}
where $\epsilon$ is an ultraviolet-cutoff length scale; its presence reflects the divergence in the partition function due to the presence of conical singularities in $\R_{1,n}$. This gives rise to the well-known formula for the EE \cite{Holzhey:1994we,cardy0},
\begin{equation}
S([u,v])=S^{(1)}([u,v]) = \frac c3\ln\left(\frac{v-u}\epsilon\right).
\end{equation}

The above result is easily generalized to the case of a single interval on a circle at zero temperature or on the line at finite temperature \cite{cardy0}. In either case, the branched cover surface continues to have genus zero, and therefore the EREs and EE continue to be proportional to $c$. The simplest cases where higher-genus partition functions appear are a single interval on the circle at finite temperature and two intervals on the line at zero temperature; the corresponding branched-cover surfaces have genera $n$ and $n-1$, respectively. This implies that the ERE will depend on the full operator content of the CFT, not just its central charge \cite{Calabrese:2009ez}. In rest of this subsection we will focus on the two-interval case, which is the best-studied one.

For two intervals $[u_1,v_1]\cup[u_2,v_2]$, it is convenient to work with the mutual (R\'enyi) information, which is ultraviolet-finite, hence conformally invariant and dependent only on the cross-ratio $y$ of the four endpoints \cite{Calabrese:2009ez}:\footnote{In the literature, this cross-ratio is often denoted $x$; however, we will use $x$ for a different cross-ratio in what follows.}
\begin{equation}
I^{(n)}(y) := S^{(n)}([u_1,v_1])+S^{(n)}([u_2,v_2])-S^{(n)}([u_1,v_1]\cup[u_2,v_2])\,,\quad
y:=\frac{(v_1-u_1)(v_2-u_2)}{(u_2-u_1)(v_2-v_1)}\,.
\end{equation}
We have
\begin{equation}\label{IFrelation}
I^{(n)}(y) = \frac1{1-n}F(\R_{2,n}) + \text{subtractions}\,.
\end{equation}
The subtractions, given by the EREs of the individual intervals, soak up the divergences in $F(\R_{2,n})$, leaving an unambiguous finite value for $I^{(n)}(y)$. The partition function on $\R_{2,n}$ can be expressed as a four-point function of twist operators in the orbifold theory:
\begin{equation}\label{twist4point}
Z(\R_{2,n}) = \ev{\sigma(u_1)\tilde\sigma(v_1)\sigma(u_2)\tilde\sigma(v_2)}_{\C^n/\Z_n}\,.
\end{equation}

The surface $\R_{2,n}$ has genus $n-1$, so the partition function depends on the full operator content of the theory and not just its central charge. However, it contains a universal contribution that only depends on $c$. To define this part, it is useful to first set up some notation regarding the topology of the surface $\R_{2,n}$. 

\begin{figure}[t!]
\centering
\includegraphics[width=0.4\textwidth]{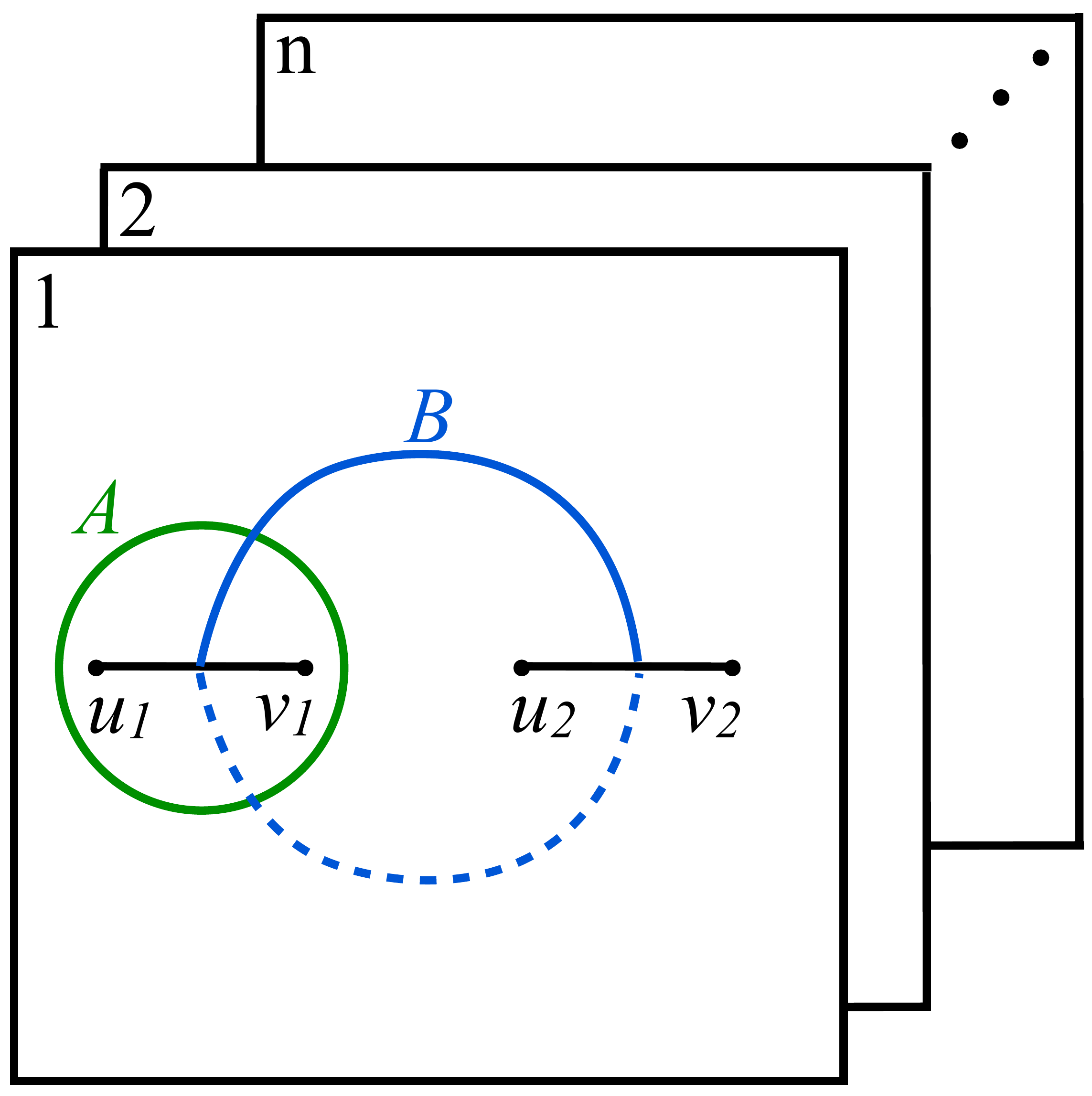}
\caption{\label{fig-R2n}The $n$-sheeted replica surface $\R_{2,n}$, which is the branched covering surface of the plane with two intervals and has genus $n-1$. On each sheet, there is a cycle separating the two intervals called the $A$-cycle, and another cycle encircling the two points $v_1$ and $u_2$, called the $B$-cycle. There are $n-1$ independent cycles of each type.}
\end{figure}

\begin{figure}[t!]
\centering
\includegraphics[width=0.6\textwidth]{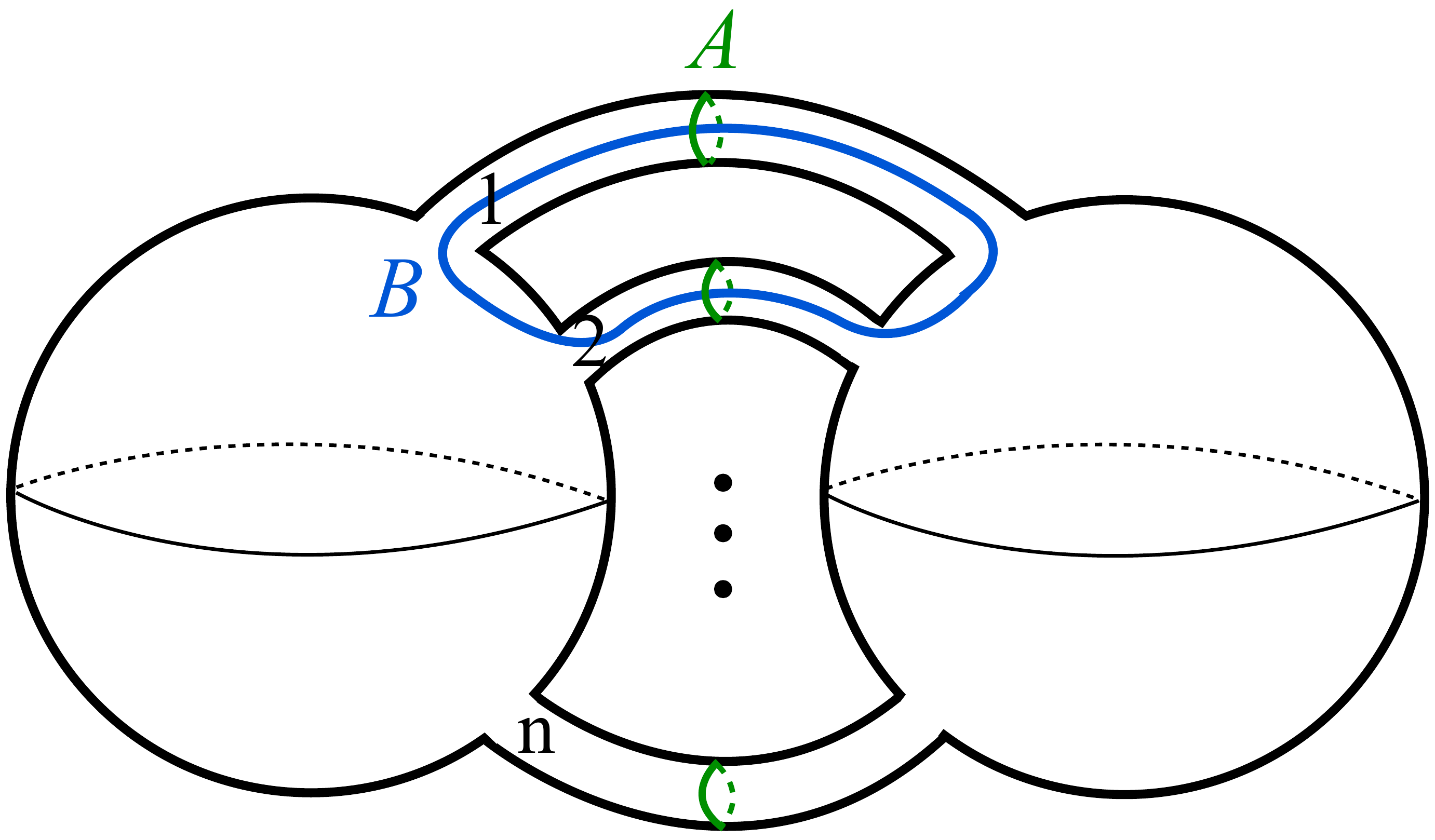}
\caption{\label{fig-spheres}An alternate depiction of the surface $\R_{2,n}$ in figure \ref{fig-R2n}. $\R_{2,n}$ can be visualized as two spheres connected by $n$ tubes. The two spheres, one for each interval, are made by cutting small holes around each pair of intervals on all $n$ sheets. The tubes connecting the holes on the two spheres represent the sheets. In this picture, the $A$-cycles wrap the $n$ tubes and the $B$-cycles run through two different tubes.}
\end{figure}

A useful basis of cycles on $\R_{2,n}$ can be described as follows. On each sheet, there is a cycle that separates the two intervals. We will call these A-cycles. The sum of all $n$ of them is trivial, so there are $n-1$ independent ones. There are also cycles which encircle the points $v_1,u_2$, crossing each cut once, which we call B-cycles; again, there are $n-1$ independent ones. (See figure \ref{fig-R2n}.) 
The A-cycles intersect the B-cycles but not themselves, and vice versa. (Linear combinations $A_i,B_j$ can be constructed with intersection numbers $A_i\cdot B_j=\delta_{ij}$, but this will not be necessary for our purposes.) It is also useful to visualize the surface $\R_{2,n}$ as two spheres connected by $n$ tubes. This can be related to the branched cover by cutting each sheet along a small ellipse surrounding the interval $[u_1,v_1]$ and another one surrounding the interval $[u_2,v_2]$. Each interval then becomes a sphere with $n$ holes, while each sheet becomes a tube connecting one sphere to the other. Each A-cycle wraps a tube, while each B-cycle runs along one tube and back along another. (See figure \ref{fig-spheres}.)
 $\R_{2,n}$ enjoys a $\Z_n$ ``replica symmetry'', which cyclically permutes the sheets, and hence also the tubes.
 
The universal part of $Z(\R_{2,n})$ to which we alluded above is defined as the contribution in which only Virasoro descendants of the vacuum appear on the A-cycles. 
In other words, if for any circle $C$ we define $P_{\rm vac}(C)$ as the projection operator onto the conformal family of the vacuum of the Hilbert space of $\C$ on $C$, then we define the \emph{vacuum partition function} as the path integral with projectors $P_{\rm vac}(A_1)\cdots P_{\rm vac}(A_{n-1})$ inserted:
\begin{equation}
Z_{\rm vac}(\R_{2,n}) := \ev{P_{\rm vac}(A_1)\cdots P_{\rm vac}(A_{n-1})}Z(\R_{2,n})\,.
\end{equation}
With only vacuum descendants on $A_1,\ldots,A_{n-1}$, the cycle $A_n$ (which is a linear combination of the others) is automatically guaranteed to host only such descendants as well.\footnote{To see this, cut along all $n$ A-cycles, leaving two sphere $n$-point functions (on the left and right spheres of figure \ref{fig-spheres}). For each $n$-point function, $n-1$ of the operators correspond to vacuum descendants. As a result, if the $n$th one is not a vacuum descendant, then the $n$-point function vanishes, and hence does not contribute to the vacuum partition function.} Note that the choice of representative of any given cycle $A_i$ is unimportant; any representative can be mapped to any other by a holomorphic diffeomorphism, which acts on the Hilbert space by the Virasoro group, under which conformal families don't mix. In the orbifold description, the vacuum partition function can be written
\begin{equation}
Z_{\rm vac}(\R_{2,n}) = \ev{\sigma(u_1)\tilde\sigma(v_1)P_{\text{orb vac}}(A)\sigma(u_2)\tilde\sigma(v_2)}_{\C^n/\Z_n}\,,
\end{equation}
where $P_{\text{orb vac}}$ is the projector onto states of $\C^n/\Z_n$ composed of descendants of the identity of $\C$,\footnote{This set of states includes more than just Virasoro descendants of the vacuum of $\C^n/\Z_n$. Rather, it includes all descendants of the vacuum under the larger algebra consisting of ($\Z_n$-symmetric) products of Virasoro generators acting on the different copies of $\C$.} and $A$ is a circle enclosing $[u_1,v_1]$. Note that, unlike the full partition function, $Z_{\rm vac}(\R_{2,n})$ is not a modular invariant quantity, due to the distinguished role of the A-cycles.

As we will see in the next subsection, the vacuum partition function is particularly well-studied in the context of holographic and other large-$c$ CFTs.

\subsubsection{Large-$c$ CFTs}

We are interested in families of CFTs, such as holographic ones, that admit a large-$c$ limit. In such theories, all of these quantities---the free energies, entanglement (R\'enyi) entropies, and mutual (R\'enyi) informations---admit an expansion in $1/c$ starting at order $c$. We thus write, for example,
\begin{equation}
I^{(n)}(y) = \sum_{\ell=0}^\infty c^{1-\ell}I^{(n)}_\ell(y)\,,\qquad
F(\R_{2,n}) = \sum_{\ell=0}^\infty c^{1-\ell}F_\ell(\R_{2,n})\,.
\end{equation}
In a holographic CFT, the parameter $1/c$ is proportional to the bulk Newton constant,
\begin{equation}
\frac1c = \frac{2G_N}{3R_{\rm AdS}}\,,
\end{equation}
so the expansion in $1/c$ is a loop expansion in the bulk (hence the index $\ell$).

From a CFT perspective, the loop corrections ($\ell\ge1$) are ``cleaner'' than the classical one ($\ell=0$), in the following sense. First, $F_{\ell\ge1}$ is unambiguous, since the scheme dependence of the free energy is due to the Weyl anomaly, which is proportional to the central charge. Second, it is finite even on a singular surface such as $\R_{2,n}$, since the Weyl transformation that smoothes out those conical singularities shifts the free energy by $c$ times a Liouville action. Third, since it is Weyl-invariant, it depends only on the complex structure of $\R_{2,n}$, hence only on the cross-ratio $y$, not the positions of the endpoints themselves. Finally, since the subtractions present in \eqref{IFrelation} are proportional to $c$, we simply have
\begin{equation}\label{IFrelation2}
I^{(n)}_\ell(y) = \frac1{1-n}F_\ell(\R_{2,n})\qquad\text{for $\ell\ge1$}\,.
\end{equation}
These properties will all be useful when we study the loop corrections below.

The RT formula makes a strikingly simple prediction for the classical part of the mutual information \cite{rt}:
\begin{equation}\label{RTpredict}
I_0(y) = \begin{cases} 0\,, & y\le1/2 \\ (1/3)\ln(y/(1-y)) & y\ge1/2 \end{cases}\,.
\end{equation}
It is interesting that this formula does not depend on the field content or other specifics of the dual theory. On the other hand, the loop corrections do depend on the field content, although they always include certain ``universal'' terms due to the gravitational sector, as we will explain below.

Significant effort has gone into testing the prediction \eqref{RTpredict} and computing the loop corrections using the replica trick. Two strategies have been followed to compute the relevant free energies. The first is to find the dominant gravitational saddle point whose conformal boundary is $\R_{2,n}$; the terms in the $1/c$ expansion of $F(\R_{2,n})$ are then given by the classical action, the one-loop determinant of the fields about that background, and so on. The second strategy is to compute the four-point function of twist fields \eqr{twist4point} using CFT techniques such as the conformal-block decomposition. The RT prediction for the classical part was successfully confirmed, modulo some assumptions, by both methods, in \cite{Faulkner:2013yia} and \cite{Hartman:2013mia} respectively.

Consider the calculation of $F_0(\R_{2,n})$, starting with the gravity method. In \cite{Faulkner:2013yia}, two gravitational saddles were constructed with conformal boundary $\R_{2,n}$. Both are handlebodies; in one, which we will call $H_A$, the A-cycles are contractible, while in the other, $H_B$, the B-cycles are contractible. $H_A$ has a smaller action for $y<1/2$ and $H_B$ for $y>1/2$. These are the only solutions that preserve the replica symmetry of $\R_{2,n}$, and are also believed to be the only type of solution that exists uniformly for all $n$. Their actions are analytic functions of $n$; when continued down to $n=1$, they reproduce precisely the RT prediction \eqref{RTpredict} for the EE. In \cite{Hartman:2013mia}, an analysis of conformal blocks in the $\C^n/\Z_n$ theory at large $c$---again, imposing the replica symmetry---led to the same result.

An important subtlety regarding these calculations is as follows. For general $n$ and $y$, it is not clear whether the dominant gravitational saddle is always $H_A$ or $H_B$, and therefore whether their actions indeed give the correct free energy and R\'enyi entropy.\footnote{Even if this is not the case, one can argue that these are the relevant saddles to consider for the purposes of analytically continuing the ERE down to $n=1$ to find the EE.} However, for small $y$, the tubes are very thin (as we will see in section \ref{v-ii} when we discuss the period matrix for $\R_{2,n}$), so the dominant saddle must indeed be the handlebody that fills them in, namely $H_A$. This is important for our purposes because the calculations we will describe from here on will always be done in an expansion in $y$, and therefore we can safely ignore this subtlety and assume that $H_A$ is the dominant saddle. 

We now turn to the one-loop correction to the free energy, $F_1(\R_{2,n})$, which as noted in \eqref{IFrelation2} directly gives the one-loop correction to the mutual R\'enyi information, $I_1^{(n)}(y)$. $F_1(\R_{2,n})$ is proportional to the sum of the logs of the fluctuation determinants of all the fields propagating on the relevant gravitational saddle. In any theory of gravity, this includes the metric fluctuations. Their one-loop determinant on the handlebody $H_A$ was computed in an expansion in $y$ to order $y^8$ for all $n$ in \cite{Barrella:2013wja}, and to order $y^{10}$ for $n=1$ in \cite{Beccaria:2014lqa}.

As we will now explain, this contribution to the free energy is simply the $\Oc(c^0)$ part of the vacuum free energy $Z_{\rm vac}(\R_{2,n})$ defined in the previous subsection. In fact, more generally, consider the partition function obtained from the classical action and loop corrections to all orders of perturbative pure gravity on $H_A$. We will now argue that this quantity is precisely $Z_{\rm vac}(\R_{2,n})$. In the genus-one case, this was shown in \cite{Maloney:2007ud}, and we can adopt their argument here. In a Hilbert-space interpretation, we can choose to think of the A-cycles as defining a spatial direction and the B-cycles a (Euclidean) time direction. This is convenient because the states defined on the A-cycles are perturbative pure quantum gravity states on an AdS${}_3$ background, since the A-cycles are contractible and the handlebody is locally Euclidean AdS$_3$. Since the creation operators for metric fluctuations are, from the CFT viewpoint, Virasoro generators, these states are thus Virasoro descendants of the vacuum. Thus the perturbative pure gravity partition function on $H_A$ is precisely $Z_{\rm vac}(\R_{2,n})$. The exact correspondence between the perturbative quantum gravity partition function and the universal identity block contributions to CFT partition functions was articulated and tested in \cite{Yin:2007gv}. We will extend that work in section \ref{v-iii}.
\vs

We now return to computation of R\'enyi entropies specifically. Having established the CFT interpretation of $Z_{\rm vac}(\R_{2,n})$, we can see that reproducing the results of \cite{Barrella:2013wja,Beccaria:2014lqa} using the twist-field method requires including only descendants of the vacuum as intermediate states in the conformal-block decomposition of the 4-point function \eqref{twist4point}, since the intermediate states are precisely those living on the A-cycles. More precisely, one should include states of the orbifold theory $\C^n/\Z_n$ that are made up of descendants of the vacuum of $\C$; these include more than just the descendants of the vacuum of $\C^n/\Z_n$. It is easy to see that the term of order $y^h$ in $I^{(n)}(y)$ is given by descendants at level $h$. 

These calculations were carried out to order $y^8$ by Chen et al. in \cite{Chen:2013dxa}. Expanding their result in powers of $1/c$, the one-loop (order $c^0$) term matched the bulk metric one-loop determinant computed earlier in \cite{Barrella:2013wja}. Their $c^{1-\ell}$ term started at order $y^{2\ell+2}$, so their results could access $\ell\leq 3$. In other words, they not only reproduced the one-loop determinant, but effectively computed two-loop and three-loop free energies, which would presumably be quite challenging from a direct bulk perturbative calculation. The coefficient at each order in $1/c$ and $y$ is a rational function of $n$. We will not reproduce these rather complicated functions here for general $n$. However, let us note the following pattern in the $n$-dependence observed by \cite{Chen:2013dxa}: 
\eq{chenpatt}{F_{\rm vac,\ell}(\R_{2,n}) =  (n-1)(n-2)\cdots (n-\ell)\left(\sum_{m=2\ell+2}^8 \alpha_{m,\ell}(n)y^m\right)+ O(y^9)~.}
The $\alpha_{m,\ell}(n)$ are functions of $n$; some of them have zeroes at positive $n$, but none of these zeroes coincide, unlike those shown in \eqr{chenpatt}. 

There are some notable features of this formula. First, $F_{\rm vac;\;\ell\ge1}(\R_{2,n})$ carries a factor of $n-1$. The fact that it vanishes at $n=1$ can be understood from the fact that the genus-zero free energy is given entirely by a Liouville action multiplied by $c$; it is also necessary, given \eqref{IFrelation2}, for the mutual R\'enyi information to have a smooth limit as $n\to1$. Second, $F_{\rm vac;\;\ell\ge2}(\R_{2,n})$ carries an overall factor of $n-2$. The fact that it vanishes at $n=2$ can be understood from the fact that the contribution of the identity family to the genus-one free energy is one-loop exact: aside from a classical (order-$c$) part, it is given by $-\ln\chi_{\rm vac}(y)$, where $\chi_{\rm vac}(y)$ is the character of the identity family, which is independent of $c$. 

Perhaps surprisingly, the $y^8$ term of $F_{\rm vac;\;3}(\R_{2,n})$ computed by Chen et al. carries an overall factor of $n-3$. (Recall that \cite{Chen:2013dxa} only computed through $O(y^8)$ in the $y$-expansion.) If the pattern \eqr{chenpatt} were to hold to {\it all} orders in $y$, this would imply a truncation in the loop expansion around handlebodies asymptotic to $\R_{2,n}$ with appropriate cycles contractible. On this basis, Chen et al.\ were led to suggest that the genus-$g$ free energy might be $g$-loop exact for all $g$, at least for the replica manifolds $\R_{N,n}$. One might even wonder whether this could be true for all genus-$g$ manifolds. One of the main purposes of this paper is to test this intriguing idea. To do this, we will calculate $Z_{\rm vac}$ on generic genus-two Riemann surfaces, $\S$, using a different technique that we describe now; this complementary approach will provide a gateway to applications to R\'enyi entropy and 3D quantum gravity. 

\subsection{Vacuum amplitudes from sewing}\label{sewing}

\begin{figure}
\centering
\includegraphics[width=0.98\textwidth]{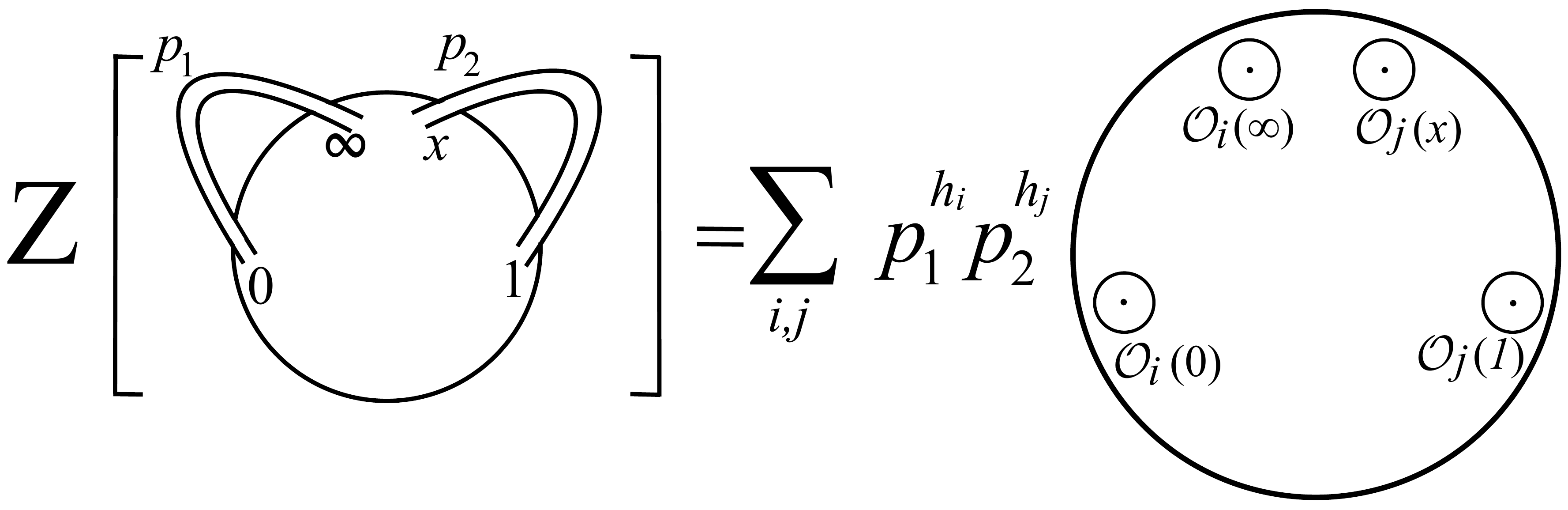}
\caption{\label{fig-sewing} A picture of the sewing approach to computing a genus-two partition function, $Z$. The mechanism was explained in figure \ref{fig-sewing-vac}. To compute $Z$ rather than $Z_{\rm vac}$, one simply lets the sum run over all operators in the CFT Hilbert space.}
\end{figure}

In section \ref{iv}, we will compute $Z_{\rm vac}$ via the sewing construction. We heuristically explain this method here with the help of figure \ref{fig-sewing}; the method applies to computation of the full partition function $Z$ of ${\cal C}$, but can be specialized to computation of $Z_{\rm vac}$. The basic idea is to replace each handle of $\S$ by a sum over local operator insertions at its ends. This frames the computation of $Z$ as a weighted sum of sphere four-point functions. As stressed earlier in this section, computing $Z_{\rm vac}$ as opposed to the full partition function of ${\cal C}$ means that we only allow Virasoro vacuum descendants to propagate along the handles. This construction is perturbative in the width of the handles. There are many parameterizations of a given surface $\S$; we use the Schottky construction, which forms $\S$ as a quotient of the Riemann sphere by a discrete subgroup of $PSL(2,\mathbb{C})$, the M\"obius group. The genus-two Schottky space is parameterized by coordinates $\lbrace p_1,p_2,x\rbrace$. Roughly speaking, these describe the width of the two handles and the sphere coordinate of the lone endpoint not fixed by conformal symmetry, respectively. The computation of $Z_{\rm vac}$ is then a double power series in $p_1$ and $p_2$, where the powers are the left-moving conformal weights of the operators inserted at the endpoints.

In order to make eventual contact with R\'enyi entropies and the work of \cite{Yin:2007gv}, we will also need to express $Z_{\rm vac}$ in terms of the period matrix of $\S$, denoted $\Om$. That is, we need to perform the coordinate map $\lbrace p_1,p_2,x\rbrace \mapsto \Om$. This is known in closed form, but is complicated (see, e.g., \cite{Yin:2007gv, Gaberdiel:2010jf}). If we define multiplicative periods 
\eq{mp}{q_{ij} := e^{2\pi i \Om_{ij}}}
then $q_{ij}$ admits a power series in $p_1$ and $p_2$ of the following form:
\es{mltprds}{
&q_{11}=p_1\sum_{n,m=0}^{\infty}p^n_1p^m_2\sum_{r=-n-m}^{n+m}c(n,m,|r|)\,x^r,\\
&q_{12}=x+x\sum_{n,m=1}^{\infty}p^n_1p^m_2\sum_{r=-n-m}^{n+m}d(n,m,r)\,x^r\\
&q_{22}=q_{11}(p_1\leftrightarrow p_2)~.
}
The $c(n,m,|r|)$ and $d(n,m,r)=d(m,n,r)$ are coefficients given in Appendix E of \cite{Gaberdiel:2010jf} through $m=n=7$. 

Thus, in order to compute $Z_{\rm vac}$ via sewing, we must compute four-point functions of operators in the Virasoro identity representation. We turn to this now, by way of the more general subject of computing four-point functions of arbitrary holomorphic operators.

\section{Four-point functions and the analytic bootstrap}\label{iii}

In this section, which may be read independently of the rest of the paper, we discuss methods for computing correlators in 2D CFTs. We will focus on four-point functions. A standard way to compute a four-point function is to do an OPE expansion of pairs of operators. This yields a power series in the cross-ratio $x$ of the four points. However, we wish to calculate the correlator at finite values of $x$. We will describe two methods to do this. The first, described in subsection \ref{iii-i}, is via direct manipulation of operator modes, and may be familiar to CFT practitioners. The second, described in subsection \ref{iii-ii}, is an analytic realization of the conformal bootstrap that applies specifically to correlators of only holomorphic (or only anti-holomorphic) operators. The upshot is that the combination of holomorphy with fundamental properties of the OPE and crossing symmetry yields a result that is determined by a {\it finite} number of OPE coefficients. In the case that all four operators have identical holomorphic dimensions, the solution of crossing symmetry leads to an especially simple algorithm. 

Before turning to those methods, we first describe a simple approach that applies specifically to correlators of descendants of the identity, and use the results to discuss the powers of the central charge $c$ that appear in such correlators.

\subsection{Free bosons and powers of $c$}\label{powers}

As explained in subsection \ref{sewing} (and illustrated in figure \ref{fig-sewing-vac}), $Z_{\rm vac}$, the universal part of the genus-two partition function, can be constructed from four-point functions on the plane of descendants of the identity. In this subsection, we will discuss general properties of such correlators. 

The first property to note is that they are independent of the rest of the field content of the theory, and depend only on its central charge. This follows from the fact that the identity Verma module is closed under fusion. Since in this paper we are particularly interested in the powers of $c$ that appear in $Z_{\rm vac}$, in this subsection we will focus on the question of what powers of $c$ can appear in such correlators. We will first use a simple counting argument in a free field theory to show that the powers of $c$ are highly constrained. In particular, if one thinks of $1/c$ as a coupling constant, then it appears that these correlators are tree-level exact. We will relate this classicality to the fact that the sphere partition function, for any CFT, has a particularly simple $c$-dependence, and then discuss its bulk interpretation for holographic CFTs. Finally, we will discuss the generalizations of these statements at higher genus.

The fact that correlators of descendants of the identity are independent of the theory, except its central charge, implies that we can compute them in any convenient theory with a variable central charge. One simple choice is the theory of $c$ free bosons; by writing the relevant operators in terms of elementary fields, it is in principle straightforward to compute their correlators using free-field Wick contractions. This procedure is fairly tractable for calculating, for example, the four-point function of the stress tensor, but it rapidly becomes unwieldy when applied to higher-point functions or higher-level descendants, and for these calculations the methods described in the following subsections are far more efficient.

Nonetheless, the free-boson method gives a fast way to answer the important question of what powers of $c$ appear in a given correlator. For example, the stress tensor is \begin{equation}
T = 
-\frac12\sum_{\mu=1}^c:\partial X^\mu\partial X^\mu:\,.
\end{equation}
An $m$-point function of stress tensors $\ev{T(z_1)\cdots T(z_m)}$ includes indices $\mu_1,\ldots,\mu_m$. The $2m$ $X$ fields appearing can be contracted in various ways, linking the different $T$'s, and therefore the different $\mu$'s, to each other. For example, a contraction between $X^{\mu_1}$ and $X^{\mu_2}$ leads to a factor of $\delta_{\mu_1\mu_2}$. In the connected part of the correlator, they are all linked in one group, so a non-zero contribution occurs only when all the $\mu$ are equal: $\mu_1=\cdots=\mu_m$. Hence the connected part of the correlator is linear in $c$, independent of $m$. Disconnected parts give higher powers of $c$; for example, the stress tensor four-point function has a term quadratic in $c$, from contractions in which the $T$ are linked in two separate pairs (see \eqref{eq8} for the explicit form).

All descendants of the identity can be written as normal-ordered products of derivatives of stress tensors. The connected part of any correlator---regardless of the number and type of operator---again comes from terms in which all the $\mu$'s are equal, and is therefore again linear in $c$. This also follows from the fact that the generating function for connected correlators of the stress tensor is the sphere free energy as a functional of the metric, which is simply $c$ times the Liouville action. Thus, if we think of $1/c$ as playing the role of $\hbar$, any CFT on the sphere is purely classical in this sector, in other words its correlators are effectively given entirely by tree-level contributions. To translate this into the usual field-theory language, if we normalize $T$ by a factor of $c^{-1/2}$ (so that its two-point function is 1), then a connected correlator with a total of $P$ factors of $T$ is proportional to $c^{1-P/2}$, just like a tree-level diagram with $P$ external legs in a field theory with coupling constant $1/c$.

We now turn to holographic theories, where $1/c\sim G_N$ is indeed the bulk coupling constant. An operator made of $p$ stress tensors corresponds to a state containing $p$ gravitons. This again leads to $c^{1-P/2}$ for a tree-level bulk process involving a total of $P$ gravitons. From this point of view, the absence of loop corrections may seem mysterious, given that there certainly exist Witten diagrams in the bulk containing loops, which make non-zero contributions to such a correlator. However, in 3D gravity, all terms in the effective action which depend only on the metric (even those generated by loops of other fields) can be absorbed in the Einstein-Hilbert term \cite{Gupta:2007th}. Hence, the full quantum effective action for the metric \emph{is} simply the classical action, with a renormalized value of the Newton constant. Since it is the renormalized Newton constant that enters in the relation $1/c=2G_N/3R_{\rm AdS}$, when working in terms of $c$, the theory appears to be entirely classical.\footnote{This property is {\it not} directly related to the absence of propagating degrees of freedom in pure 3D gravity. To demonstrate this, one can consider the correlator of four spin-$s$ currents with $s>2$. As we will show by example later in section \ref{identop} (see \eqr{Lamb}--\eqr{wcoeffs}), these correlators do not truncate in a $1/c$ expansion. This implies a non-trivial loop expansion for bulk four-point scattering of spin-$s$ gauge fields in pure 3D higher spin gravity, even though these theories also lack propagating modes.}

The arguments above, both in the field theory and in the bulk, depend crucially on the fact that we are working on the plane (or, more generally, on the sphere with any metric). On surfaces with non-zero genus, as we recalled in subsection \ref{Renyireview}, the free energy (or effective action) includes higher powers of $1/c$, and these higher-order terms depend on the full operator content of the theory. So one would not expect correlators to be purely classical. Similarly, from a bulk point of view, gravitons and other particles can propagate in loops that wrap non-trivial cycles of the bulk, giving corrections that cannot be captured by a local effective action.

Nonetheless, as noted below \eqref{chenpatt}, at genus one, the \emph{vacuum} free energy $F_{\rm vac}(T^2)=-\ln Z_{\rm vac}(T^2)$ does have the special property that it is one-loop exact, in other words contains only terms linear and constant in $c$ (in any CFT). $Z_{\rm vac}(T^2)$ is defined as the path integral with the insertion of the operator $P_{\rm vac}(A)$ that projects onto vacuum descendants on some fundamental cycle $A$ of the torus. Derivatives of the free energy with respect to the metric give connected correlators of the form $\ev{P_{\rm vac}(A)\mathcal{O}_1\mathcal{O}_2\cdots}_{\rm con}$, where the $\mathcal{O}_i$ are descendants of the identity. Such correlators are therefore also one-loop exact (contain only terms linear and constant in $c$). We will confirm this property by explicit calculation in subsection \ref{iv-ii} below.

\subsec{Operator modes}\label{iii-i}

To begin, we will compute vacuum correlators of the form 
\eq{eq1}{\langle \Oc(\infty) T(1) T(z) \Oc(0) \rangle}
where $\langle \cdot \rangle := \langle 0|\cdot|0\rangle$. The operator $\Oc$ is allowed to be an arbitrary, non-holomorphic operator, not necessarily primary or quasi-primary. As is conventional, we leave its anti-holomorphic dependence implicit in what follows. We define mode expansions
\eq{eq2}{T(z) = \sum_{n\, \in \, \mathbb{Z}} {L_n\over z^{n+2}}~, \quad \Oc(z) = \sum_{n\,\in\, \mathbb{Z}} {\Oc_{n}\over z^{n+h}}}
where the stress tensor modes obey the Virasoro algebra,
\eq{eq3}{[L_m,L_n] = (m-n) L_{m+n} + {c\over 12}n(n^2-1) \delta_{m+n,0}~.}
In terms of modes, the four-point function is
\eq{eq5}{\langle \Oc(\infty) T(1) T(z) \Oc(0) \rangle = z^{-2} \sum_{n\, \in \, \mathbb{Z}} z^{-n} \langle \Oc_{h} L_{-n} L_n \Oc_{-h}\rangle~.}

To proceed, we break up the sum into the $n=0$ mode term, and two sums over positive and negative integers (denoted $\Z_+$ and $\Z_-$, respectively). Using the fact that $L_0\Oc_{-h} |0\rangle = h\Oc_{-h}|0\rangle$, the $n=0$ mode contributes a term $z^{-2} h^2 \N_{\Oc}$, where $\N_{\Oc} = \langle \Oc_{h} \Oc_{-h}\rangle$ is the norm of $\Oc$. Using the Virasoro algebra, and relabeling $n\rar -n$, we can rewrite the sum over $\Z_-$ in terms of a sum over $\Z_+$ as
\bea\label{eq7}
\sum_{n\, \in \, \mathbb{Z}_-}  z^{-n} \langle \Oc_{h} L_{-n} L_n \Oc_{-h}\rangle \!\!\!&=&\!\!\! \sum_{n\, \in \, \mathbb{Z}_+}z^{n}\Big(\langle \Oc_{h} L_{-n} L_{n} \Oc_{-h}\rangle+\left(2nh+{c\over 12}n(n^2-1)\right) \N_{\Oc} \Big)\quad \quad \quad\\
\!\!\!&=&\!\!\! \sum_{n\, \in \, \mathbb{Z}_+}z^n  \langle \Oc_{h} L_{-n} L_{n} \Oc_{-h}\rangle+\left({c\over 2}{z^2\over (1-z)^4}+2h{z\over (1-z)^2}\right)\N_{\Oc}~.\nonumber
\eea
The quantity in angle brackets is simply the squared norm $||L_n \Oc_{-h}|0\rangle||^2$. 

There is a further simplification of this sum: it truncates on account of vacuum invariance. Suppose $\Oc$ is a level $N$ descendant of a primary field $\Oc'$ of holomorphic dimension $H=h-N$. Then $\Oc_{-h}$ can be written as a linear combination of ``lexicographically ordered'' operators,
\eq{eq9}{L_{-n_1}\cdots L_{-n_k} \Oc'_{-H}}
where $n_1 \geq n_2 \geq \ldots \geq n_k$ and $N = \sum_{i=1}^k n_i$. This implies that the sum in \eqr{eq7} truncates at $n=N$, because $L_n \Oc_{-h}|0\rangle = 0$ for $n>N$ by definition of a primary. 

Taking this into account and adding \eqr{eq7} to the other pieces, the full correlator is
\es{eq8}{&\langle \Oc(\infty) T(1) T(z) \Oc(0) \rangle = \\&z^{-2}\left(\sum_{n=1}^N(z^n+z^{-n}) \langle \Oc_{h} L_{-n} L_{n} \Oc_{-h}\rangle +\left({c\over 2}{z^2\over (1-z)^4}+2h{z\over (1-z)^2}+h^2\right)\N_{\Oc}\right)~.}
A pleasing feature of the expression in parenthesis is its manifest invariance under $z\rar 1/z$, which is simply invariance under crossing symmetry corresponding to exchange of the two stress tensors. 

If $\Oc$ is quasi-primary, then the $n=1$ term of the sum vanishes. If $\Oc$ is primary, the entire sum vanishes. That leaves us with a very simple expression:
\es{eq10}{\Oc~\text{primary}:\quad \langle \Oc(\infty) T(1) T(z) \Oc(0) \rangle =z^{-2}\left({c\over 2}{z^2\over (1-z)^4}+2h{z\over (1-z)^2}+h^2\right)\N_{\Oc}~.}

A useful check of \eqr{eq8} is to take $\Oc=T$, which yields the stress tensor four-point function. Using the Virasoro algebra to compute $\langle L_2 L_{-2} L_2 L_{-2} \rangle = c^2/4$, we find
\es{TTTT}{\langle T(\infty)T(1) T(z) T(0) \rangle &=z^{-4}\left({c^2\over 4}\left(1+z^4+{z^4\over (1-z)^4}\right) + 2c{(1-z+z^2)\over (1-z)^2}\right)~.}
This agrees with previous results (e.g. \cite{Osborn:2012vt}). We may define the conformal cross-ratio as
\eq{eq4}{x := {z_{12}z_{34}\over z_{13}z_{24}} = z~,}
in which case \eqr{TTTT} has the correct form of a four-point function as determined by conformal symmetry: %
\eq{TTTT2}{\langle T(\infty)T(1) T(z) T(0) \rangle = z^{-4} \cF(x)~.}
We will reproduce this result in the next subsection in a more efficient way.\footnote{Following our discussion in subsection \ref{powers}, note that this could be derived for {\it all} $c$ from a 3D gravity computation at {\it large} $c$, by thinking of $T$ as a single-graviton state: the $\Oc(c^2)$ part is the free-field Wick contraction, and the $\Oc(c)$ part is the connected bulk correlator of four gravitons expressed in terms of the renormalized Newton constant.}

One can easily generalize this analysis to correlators where $T$ is replaced by a different operator. For simplicity, we consider the four-point function of two pairs of holomorphic quasi-primaries $\Oc^a$ and $\Oc^b$, of dimensions $h_a$ and $h_b$, respectively. (Their modes are defined as in \eqr{eq2}.) In this case, the resulting expression is
\bea\label{oooo}
&&\!\!\!\!\!\!\!\langle \Oc^a(\infty)\Oc^b(1)\Oc^b(z)\Oc^a(0)\rangle=\\
&&\!\!\!\!\!\!\!z^{-h_j}\left(\sum_{n=1}^{h_a} (z^n+z^{-n})\langle \Oc^a_{h_a}\Oc^b_{-n}\Oc^b_n\Oc^a_{-h_a}\rangle + \langle \Oc^a_{h_a}\Oc^b_0\Oc_0^b\Oc^a_{-h_a}\rangle+ \sum_{n\, \in \, \mathbb{Z}_+}z^n\langle \Oc^a_{h_a} [\Oc^b_n,\Oc^b_{-n}]\Oc^a_{-h_a}\rangle\right)\nonumber
\eea
To understand why the first sum truncates at $n=h_a$, we need to examine the OPE between quasi-primaries: in terms of modes,
\eq{qpope}{[\Oc^a_m,\Oc^b_n] = \sum_c C^{ab}_{~~c} P(m,n;h_{a},h_{b}; h_{c})\, \Oc^c_{m+n} + G^{ab}\delta_{m+n,0} \left({\begin{array}{c}
m+h_a-1  \\
h_a+h_b-1  \\
\end{array} } \right) ~.}
$G^{ab}$ is the Zamolodchikov metric, $ \Oc^c$ are also quasi-primary, $C^{ab}_{~~c}$ are OPE coefficients, and the $P(m,n;h_{a},h_{b}; h_{c})$ are known functions\footnote{See e.g. equation (3.4) of \cite{Bowcock:1990ku}. All $P(m,n;h_{a},h_{b}; h_{c})$ are finite in unitary CFTs for operators of finite dimension.} encoding the contribution of the full global conformal family of $\Oc^c$. All modes $\Oc^b_{n}$ with $n> -h_b$ annihilate the vacuum. This enables us to write $\Oc^b_n\Oc^a_{-h_a}|0\rangle = [\Oc^b_n,\Oc^a_{-h_a}]|0\rangle$ for $n>-h_b$; the OPE \eqr{qpope}, combined with the unitarity bound $h\geq 0$, ensures that modes with $n>h_a$ give vanishing contribution. This explains the upper bound in \eqr{oooo}.

Equation \eqr{oooo}, while compact, is not particularly transparent. Even if $\Oc^b$ is made of current modes alone, its modes may be given by infinite sums over products of the $L_n$, which are difficult to manipulate. More generally, the four-point function appears to depend on the full holomorphic operator content of the theory, due to the presence of the commutator $[\Oc^b_n, \Oc^b_{-n}]$. In fact, this latter point belies the true structure of the result. We now demonstrate this explicitly as we turn to a much more powerful method of computation for correlators of holomorphic operators.

\subsec{The holomorphic bootstrap}\label{iii-ii}

We will now describe a general method to compute the correlation functions of chiral operators using crossing symmetry.
We will see that any correlation function of chiral operators which obey a closed operator product algebra may be determined uniquely by a {\it finite} number of three-point function coefficients.  This is in contrast to the typical situation, where the OPE allows us to determine correlation functions only in terms of an infinite sum over intermediate states.
In many cases, such as for the correlation functions of Virasoro descendants of the identity,  this leads to an extremely efficient computational algorithm.

Let us recapitulate our conventions for chiral operators. We make no further reference to the mode notation of the previous subsection. We will consider a family of chiral operators $\Oc_a(z)$, with integer dimensions $h_a$, and ${\bar h}_a=0$. We will take the basis $\Oc_a$ to be quasi-primaries and assume that the $\Oc_a$ satisfy a closed OPE
\be
\Oc_a(z_1) \Oc_b(z_2) \sim \sum_c C_{ab}{}^c {\Oc_c (z_2)\over z_{12}^{h_a+h_b-h_c}}~+~(\text{descendants})
\ee
The two point functions
\be
\langle \Oc_a (z_1) \Oc_b (z_2) \rangle = {G_{ab} \over z_{12}^{h_a+h_b}} 
\ee
and three-point functions
\be
\langle \Oc_a(z_1) \Oc_b(z_2) \Oc_c(z_3) \rangle = {C_{abc} \over z_{12}^{h_a+h_b-h_c}z_{13}^{h_a+h_c-h_b}z_{23}^{h_b+h_c-h_a}}
\ee
are fixed, up to constants, by conformal invariance.  

\subsubsection{Four-point functions: General case}

Conformal invariance constrains the four-point function to take the form
\be
\label{4pt}
\langle \Oc_a(z_1) \Oc_b(z_2) \Oc_c(z_3) \Oc_d (z_4) \rangle 
=\left({ 1\over z_{12}^{h_a+h_b} z_{34}^{h_c+h_d}} \left({z_{24} \over z_{14}} \right)^{h_{ab} } \left({z_{14} \over z_{13}}\right)^{h_{cd}} \right){\cal F}_{abcd}(x)  ~.
\ee
where we define the cross ratio $x$ as in \eqr{eq4},
\be
x={z_{12} z_{34} \over z_{13} z_{24}}~, ~~~~~~~~~~~~~~~~ 1-x={z_{14} z_{23} \over z_{13} z_{24}} ~.
\ee
We will use the notation $H=\sum_a h_a$, $h_{ab}=h_a-h_b$, $z_{ab}=z_a-z_b$, etc.

Our starting point is the observation  that the four-point function (\ref{4pt}) depends analytically on the $z_i$ and has poles only when the points $z_i$ coincide.  Thus ${\cal F}_{abcd}$ is a meromorphic function of $x$ with poles only at $x=0,1,\infty$.   So ${\cal F}_{abcd}$ is a rational function of $x$, which is uniquely completely determined (up to a constant piece) by its polar behaviour at these points.  As we will see, this polar behaviour is fixed by only a finite number of three-point function coefficients.

We begin by considering the expansion of $\cF_{abcd}$ near $x\to0$. 
This can be found by inserting the $\Oc_a \Oc_b$ and $\Oc_c \Oc_d$ OPE into the four-point function (\ref{4pt}).  The result is a sum over intermediate operators $\Oc_e$.  The contributions from the descendant states of a given quasi-primary are given by a rigid (i.e. $SL(2,\mathbb{R}$)) conformal block.  The rigid conformal blocks were written in terms of hypergeometric functions in  \cite{zamo}.  The result is
\be
\label{cblock}
{\cal F}_{abcd} (x)= 
\sum_{e} \left(C_{abe} C_{cd}{}^e \right)
x^{h_e}F(h_e-h_{ab}, h_e+h_{cd};2h_e;x)
\ee
From this we see that ${\cal F}_{abcd}$ is finite as $x\to 0$.  The constant term as $x\to0$ is given by the exchange of the identity operator, so 
\be
{\cal F}_{abcd} (x) = G_{ab} G_{cd} + \dots \label{0exp}
\ee
where $\dots$ denotes terms that vanish as $x\to 0$.

We now need to determine the polar behaviour near $x\to1$ and $x\to\infty$.  To do this we will use the transformation properties of
the four-point function under crossing symmetry.  The crossing symmetry conditions can be derived by considering how the correlation function (\ref{4pt}) transforms when the $z_i$ are permuted.  In particular, let us consider a permutation $\pi\in S_4$ of four elements.  We have 
\be
\langle \Oc_a(z_1) \Oc_b(z_2) \Oc_c(z_3) \Oc_d (z_4) \rangle =
\langle O_{\pi(a)}(z_{\pi(1)}) O_{\pi(b)}(z_{\pi(2)}) O_{\pi(c)}(z_{\pi(3)}) O_{\pi(d)} (z_{\pi(4)}) \rangle
\ee
This relates ${\cal F}_{abcd} (x)$ to ${\cal F}_{\pi(abcd)} (\pi (x))$, where the permutation $\pi$ acts on the cross-ratio as
\be
\pi(x) \equiv {z_{\pi(13)} z_{\pi(24)} \over z_{\pi(12)} z_{\pi(34)}}
\ee
One just needs to determine how the permutation $\pi$ acts on the prefactor in parenthesis in equation (\ref{4pt}). 
Some permutations have $\pi(x)=x$; these give identities for the ${\cal F}_{abcd}(x)$ with fixed $x$.  One can verify that these identities follow immediately from the conformal block expansion (\ref{cblock}), using properties of the hypergeometric function identities and symmetries of the three-point function coefficients.
Other permutations act on $x$, and give non-trivial information about four-point functions. In particular, the permutations $\pi=(14)$ and  $\pi=(24)$ give the crossing symmetry equations
\es{cros}{{\cal F}_{abcd}(x) &= (-1)^{H} x^{h_a + h_d} {\cal F}_{dbca}({1/x})\\
&= (-1)^{H} x^{h_c + h_d} (1-x)^{-h_c-h_b}   {\cal F}_{adcb}({1-x})}
These crossing equations strongly constrain the allowed form of the three-point function coefficients.
Since ${\cal F}_{abcd}(x)$ is finite as $x\to0$, we see that ${\cal F}_{abcd}$ has a pole of order $h_a+h_d$ at $x\to \infty$ and a pole of order $h_b+h_c$ at $x\to1$. 

We need to understand better the behaviour near these poles.
To determine the behaviour near $x\to\infty$ we insert the conformal block expansion (\ref{cblock}) into the first crossing symmetry equation to get
\bea
{\cal F}_{abcd}(x) &=& (-1)^{H} \sum_{e} C_{dbe} C_{ca}{}^e x^{h_a + h_d-h_e} {F}(h_e-h_{db},h_e+h_{ca};2h_e;{1/x})
\\
&\sim& \sum_{n=1}^{h_a+h_d} \alpha_n x^{n} + \dots~~~~~~~{\rm as}~x\to\infty ~.
\label{inftyexp}
\eea
Here $\dots$ denotes terms that are finite at $x\to\infty$.
The important point is that, because the hypergeometric function is finite as $x\to\infty$, the only terms that contribute to the pole are those with $h_e<h_a+h_d$.
In particular, the power series expansion of the hypergeometric function at $x\to\infty$ gives an explicit formula for the $\alpha_n$ in terms of the three-point function coefficients $C_{dbe} C_{ca}^e$ with $h_e<h_a+h_d$.  We find
\be
\alpha_n = (-1)^H \sum_{h_e=0}^{h_a+h_d-n} C_{dbe} C_{ca}{}^e{ (h_e-h_{db})_{h_a+h_d-h_e-n} (h_e+h_{ca})_{h_a+h_d-h_e-n} \over(h_a+h_d-h_e-n)! (2h_e)_{h_a+h_d-h_e-n}}~.
\ee
Similarly, near $x\to1$ we have
\es{1exp}{{\cal F}_{abcd}(x) &= (-1)^{H} \sum_{e} C_{ade} C_{cb}{}^e (1-x)^{h_e-h_c-h_b} {F}(h_e-h_{ad},h_e+h_{cb};2h_e;{1-x}) x^{h_c + h_d}
\\
&\sim \sum_{n=1}^{h_b+h_c} \beta_n (1-x)^{-n} + \dots~~~~~~~{\rm as}~x\to1 ~.}
where $\dots$ denotes terms that are finite as $x\to1$.
Again, the hypergeometric function has a simple power series expansion at $x\to1$, giving an explicit formulas for the coefficients $\beta_n$ in terms of the three-point function coefficients with $h_e<h_b+h_c$.  The formula for the $\beta_n$ is a bit more complicated than that for $\alpha_n$, since we must expand $x^{h_c + h_d}$ in powers of $1-x$ as well as the hypergeometric function, so we will not write it explicitly.  However, the important point is that there is a completely explicit (albeit complicated) expression for the $\beta_n$ in terms of the three-point function coefficients $C_{ade} C_{cb}{}^e$ with $h_e < h_b+h_c$.

The four-point function ${\cal F}_{abcd}$ is now completely fixed.  It is the unique rational function of $x$ which is finite everywhere except at $1$ and $\infty$, where its polar behaviour given by (\ref{inftyexp}) and (\ref{1exp}), and whose value at $x=0$ is given by (\ref{0exp}):
\be
\label{Fis}
{\cal F}_{abcd}(x) =  G_{ab}G_{cd}+\sum_{n=1}^{h_a+h_d} \alpha_n  x^{n} + \sum_{n=1}^{h_b+h_c} \beta_n \left[ \left({1-x}\right)^{-n} -1\right]~.
\ee
We see that ${\cal F}_{abcd}$ depends on a total of $H = h_a+h_b+h_c+h_d$ coefficients, $\alpha_n, \beta_n$, which are determined by combinations of a finite number of three-point function coefficients.  
This is a consequence of crossing symmetry applied in a holomorphic setting; for non-holomorphic operators, there is no simple formula for a four-point function in terms of a finite number of operators.

This has a remarkable consequence for the conformal bootstrap program, where crossing symmetry is used to place constraints on the three-point function coefficients. This is especially true for chiral CFTs. In a typical CFT, the bootstrap results in equations involving an infinite number of three-point function coefficients, which can only be solved by truncating or approximating the crossing symmetry equations in some way.  For a chiral CFT, the constraints are all written in terms of a finite number of equations. For example, by comparing \eqr{cblock}  with the expansion of \eqr{Fis} around $x=0$ we can obtain explicit formulas for all of the coefficients $C_{abe} C_{cd}{}^e$, for all $e$, in terms of the coefficients $C_{dbe} C_{ca}{}^e$ with $h_e<h_a+h_d$ and $C_{ade} C_{cb}{}^e$ with $h_e<h_b+h_c$. Of course, our results also apply to chiral operators in non-chiral CFTs.  
 
Moreover, we note that (\ref{Fis}) is not the unique way of writing the four-point function.  In writing (\ref{4pt}) we chose to separate out a particular combination of $z_{ij}$ to define a meromorphic function.  This choice led to a meromorphic function depending on $H$ coefficients which were determined by three-point function coefficients $C_{dbe} C_{ca}{}^e$ with $h_e<h_a+h_d$ and $C_{ade} C_{cb}{}^e$ with $h_e<h_b+h_c$.
Other ways of separating out a meromorphic function will lead to different expressions which in some cases may be more useful.  For example one particular interesting way of imposing the crossing symmetry relations is to write the four-point function as 
\bea\label{4ptalt}
{F_{abcd}(x)
\over 
z_{12}^{h_a +h_b-H/3}
z_{13}^{h_a +h_c-H/3}
z_{14}^{h_a +h_d-H/3}
z_{23}^{h_b +h_c-H/3}
z_{24}^{h_b +h_d-H/3}
z_{34}^{h_c +h_d-H/3}
}~
\eea
where 
\be
 {\cal F}_{abcd} (x) = 
 x^{H/3} (1-x)^{h/3-h_b-h_c} F_{abcd}(x)
\ee
The function $F_{abcd}$ is convenient because it treats the four points democratically -- which  makes the crossing symmetry equations very simple -- but does so at the price of introducing a branch cut in $F_{abcd}$ coming from the fractional powers of $H/3$. $F_{abcd}$ has singularities of order ${H/3}$ at each of the three points $x=0,1,\infty$; the crossing equations determine $F_{abcd}$ to be 
\be
F_{abcd}(x) = \sum_{n=0}^{\left\lfloor H/3\right\rfloor} \left(a_n x^{n-H/3} + 
 b_n x^{H/3-n}+
 c_n (1-x)^{n-H/3}
 \right)
\ee
where the $a_n, b_n, c_n$ are determined by the three-point functions of operators with $h_e\leq{\left\lfloor H/3\right\rfloor}$.  Note that, since $n=0, \dots, {\left\lfloor H/3\right\rfloor}$ we now have $3{\left\lfloor H/3+1\right\rfloor}$ coefficients to determine. It is reasonably straightforward, though tedious, to write explicit expressions for these coefficients in terms of the three-point functions.  The advantage of this approach is that it will, in principle, require the computation of fewer three-point function coefficients.  For example, if the number of operators in the chiral algebra increases rapidly with dimension (as in the case of the Virasoro algebra) then this expansion would be much more efficient.  

\subsubsection{Four-point functions: Identical operators}\label{identop}

Let us now simplify to the case where the four operators $\Oc_a$ are identical operators $\Oc$ of weight $h$, which is of interest for our computation of the higher genus partition function. In this case  the above procedure simplifies considerably.  The four-point function ${\cal F}(x) = {\cal F}_{abcd}(x)$ is a meromorphic function with poles only at $x=1,\infty$ which obeys the simplified crossing symmetry equation
\be\label{crossingsimple}
{\cal F}(x) = x^{2h} {\cal F}(1/x) = x^{2h} (1-x)^{2h} {\cal F}(1-x)
\ee
In fact, the space of such functions is a vector space of dimension $1+\lfloor2h/3\rfloor$.  
To see this, consider the function 
\be
a(x) = {(1-x+x^2)^2 \over (1-x)^2}
\ee which obeys (\ref{crossingsimple}) with $h=1$.  The function ${\cal F}(x) a(x)^{-h}$ is invariant under the anharmonic group generated by $x\to 1-x$ and $x\to 1/x$.  Moreover, this function is analytic everywhere on the Riemann sphere with the exception of a pole of order $2h$ at $x = \e^{\pi i/3}$, along with a mirror image pole at $x=\e^{-\pi i /3}$.  These points are order-three fixed points of the anharmonic group, so when expanded around $x=\pm e^{\pi i /3}$, only cubic powers may appear.
The function
\be\label{k}
k(x) = {x^2 (1-x)^2\over (1-x+x^2)^3} 
\ee
is the unique meromorphic function invariant under the anharmonic group that has a pole of order 3 at $x = \pm \e^{\pi i/3}$.  
We can therefore expand
${\cal F}(x) a(x)^{-h}$ in integer powers of $k$, to obtain
\es{F(x)}{{\cal F}(x) 
&= \sum_{n=0}^{\lfloor2h/3\rfloor} c_n {x^{2n} (1-x+x^2)^{2h-3 n}\over (1-x)^{2h-2n} }}
To implement this way of computing four-point functions, we note that to determine the coefficients $c_n$, we now simply expand this function in powers of $x$ and use the OPE to determine these coefficients as products of three-point functions. 

It is instructive to phrase our conclusions in the language of modular functions. Equating $x$ with the modular lambda function
\eq{mod1}{x = \l(\t) \approx 16 q^{1/2} - 128 q + O(q^{3/2})~,}
where $q=e^{2\pi i \tau}$, gives a map from ${\cal M}_{0,4}$ (the moduli space of four marked points on the sphere) to ${\cal M}_{1,0}$ (the moduli space of a torus). Accordingly, $SL(2,\mathbb{Z})$ transformations of $\t$ induce anharmonic group transformations of $x$: specifically, $x\rar 1-x$ and $x \rar 1/x$ are induced by the $S$ and $TST$ transformations, respectively. The problem of finding a function invariant under the anharmonic group therefore maps to finding a modular function, with desired polar structure in $q$ determined by the poles in $x$ via \eqr{mod1}. Given the identification \eqr{mod1}, our function $k(x)$ in \eqr{k} is just (256 times) the inverse of the $J$ function: $k(x) = 256/J(\t)$. So the construction of the four-point function ${\cal F}(x)$ is literally identical to that of torus partition functions of holomorphic CFTs, as in \cite{Witten:2007kt}. Likewise, \eqr{F(x)} implies a Rademacher expansion for OPE coefficients of higher dimension operators. 

We now treat some useful examples. For $h=1$ we have 
\be
{\cal F}(x) = c_0 {(1-x+x^2)^2\over(1-x)^2}
\ee
This is exactly the four-point function of a spin-1 current, $j$.  The coefficient $c_0=k^2$ is determined by the first (trivial) OPE coefficient $j j 1$, where $k$ is the level of the current algebra.

For $h=2$ we get two possible functions,
\be
{\cal F}(x) = { c_0 (1-x+x^2)^4 + c_1 (1-x)^2 x^2  (1-x+x^2) \over (1-x)^4}
\ee
This is the stress tensor four-point function. Matching the small $x$ expansion with OPE coefficients of $T T 1$ and $T T T$, we find $c_0= c^2/4$ and $c_1 = c(2-c)$, where we used the canonical norm for the stress tensor, $\N_T = c/2$ . This matches $\langle TTTT\rangle$ as computed in \eqr{TTTT}. 

For $h=3$ we have
\be\label{F3}
{\cal F}(x) = { c_0 (1-x+x^2)^6 + c_1 (1-x)^2 x^2  (1-x+x^2)^3 + c_2 (1-x)^4 x^4 \over (1-x)^6}
\ee
This is the four-point function of a spin-3 current, call it $W$, which was first worked out in
\cite{Zamolodchikov:1985wn} in the context of CFTs with ${\cal W}_3$ symmetry.\footnote{We believe our result actually corrects a sign error in \cite{Zamolodchikov:1985wn}: the parameter $\mu$ there should be a sum, not a difference, of two terms.} Matching the small $x$ expansion with OPE coefficients determines the $c_i$. In a theory with ${\cal W}_3$ symmetry, the first three quasi-primary operators appearing in the exchange channel of $\langle WWWW\rangle$ are $1,T$ and the level-four quasi-primary $\L$, which is the normal-ordered product of $T$ with a derivative term subtracted:
\be\label{Lamb}
\L := (TT) - {3\over 10}\p^2 T~.
\ee
This operator has norm $\N_{\L} = c(5c+22)/10$. The OPE coefficient ${WWW}$ vanishes, as it does for $W$ being any odd-spin chiral quasi-primary. Computing OPE coefficients using the Virasoro algebra and matching to the small-$x$ expansion of \eqr{F3}, we find
\eq{wcoeffs}{c_0 = \N_W^2~, \quad c_1 = \N_W^2\left({6(3-c)\over c}\right)~, \quad c_2 = \N_W^2\left( {3(5c^2-71c-102)\over c(5c+22)}\right)}
where $\langle W_3W_{-3}\rangle \equiv \N_W\propto c$ is the norm of $W$.\footnote{If the chiral algebra contains a spin-4 current too, as in the case of the ${\cal W}_{\infty}[\lambda]$ algebra appearing in the context of higher spin AdS/CFT \cite{Gaberdiel:2011wb}, this current will also appear at level four with nonzero OPE coefficient, and will change the value of $c_2$. This generalization is simple to compute using the ${\cal W}_{\infty}[\lambda]$ algebra; one instead finds $c_2 = 3\N_W^2(\lambda^2(25c^2-115c+546)-100c^2-740c-7464)/(5c(5c+22)(\lambda^2-4))$. This agrees with a previous result \cite{Long:2014oxa}.} As explained in section \ref{powers}, $\langle WWWW\rangle$ does not truncate in a $1/c$ expansion.

For $h=4$ we get 
\be\label{F4h}
{\cal F}(x) = { c_0 (1-x+x^2)^8 + c_1 (1-x)^2 x^2 (1-x+x^2)^5 + c_2 (1-x)^4 x^4 (1-x+x^2)^2 \over (1-x)^8}
\ee
An example of an $h=4$ chiral operator whose four-point function we will need in the sewing construction of $Z_{\rm vac}$ is the quasi-primary $\L$ introduced in \eqr{Lamb}. Again computing the $c_i$ by matching to OPE coefficients, we find
\eq{348}{c_0 = \N_{\L}^2~, \quad c_1 = \N_{\L}^2\left( \frac{32}{c}-8\right)~,\quad c_2 = \N_{\L}^2\left( \frac{4 \left(125 c^2+590 c+3704\right)}{5 c (5 c+22)}\right)}

In closing, one interesting comment is that certain general consequences can be immediately read off from \eqr{F(x)}. For example, when $h$ is not a multiple of 3 every contribution to the four-point function ${\cal F}(x)$ includes a factor of $1-x+x^2$.  Thus if $h$ is not a multiple of 3, the  four-point function  vanishes when $x = \pm \e^{\pi i/3}$.


\section{Genus-two partition functions}\label{iv}
We are now ready to compute $Z_{\rm vac}$ at genus two, which captures the contribution of the Virasoro vacuum module to the partition function of a CFT on a genus-two Riemann surface. We begin this section by reviewing the Schottky uniformization of generic genus-$g$ Riemann surfaces and the sewing construction of the genus-$g$ partition function. We then turn to the actual computation of $Z_{\rm vac}$ at genus two using sphere four-point functions of low-lying Virasoro vacuum descendants. The final result can be found by substituting the results of subsection \ref{iv-ii} into equation \eqr{Z2}. We focus on the holomorphic part of $Z_{\rm vac}$ henceforth.

\subsection{Schottky uniformization and the partition function}\label{iv-i}

A non-singular genus-$g$ Riemann surface can be constructed by cutting out $2g$ disks on the Riemann sphere and identifying pairs of boundary circles to form $g$ handles. The Schottky uniformization of the Riemann surface entails the identification of pairs of circles through M\"obius transformations $\gamma_i$, $i\in\{1,2,\cdots,g\}$, which are elements of $PSL(2,\bb C)$. The maps $\gamma_i$ form the generators of the Schottky group, $\Gamma$. There are three parameters $\{a_i,r_i,p_i\}$ associated with the $i^{\rm{th}}$  handle: $(a_i,r_i)$ are the locations of the centers of the boundary circles, and $p_i$ determine the width of the handles. A global conformal transformation can fix the positions of three boundary circles on the sphere and thus a genus-$g$ Riemann surface has $3g-3$ complex moduli. 

We consider the Schottky uniformization of a genus-$g$ Riemann surface following the conventions of \cite{Gaberdiel:2010jf} (see their Appendix C), where the locations of each pair of identified circles are given by the M\"obius transformation
\be
\gamma_{a_i,r_i}(z)=\frac{r_iz+a_i}{z+1},\label{mobius-i}
\ee
where $\gamma_{a_i,r_i}(0)=a_i$ and $\gamma_{a_i,r_i}(\infty)=r_i$. The generators of the Schottky group are given in terms of this map as
\be\label{sch-gens}
\gamma_i=\gamma_{a_i,r_i}\gamma_{p_i}\gamma_{a_i,r_i}^{-1},
\ee
where $\gamma_{p_i}(z)=p_iz$. We note that identified circles have opposite orientations: for the $i^{\rm{th}}$ pair the two boundary circles are given by the maps $C_i=\gamma_{a_i,r_i}\gamma_{R_i}C$ and $\bar C_{-i}=\gamma_{a_i,r_i}\hat\gamma\gamma_{R_{-i}}C$, where $C$ is the unit circle at the origin, $R_i$ and $R_{-i}$ are the radii of $C_i$ and $\bar C_{-i}$ respectively, and $\hat\gamma$ is the inverse map
\be\label{gamma1z}
\hat\gamma(z)=\frac1z.
\ee
The product of the radii of the two circles is $R_i\,R_{-i}=p_i$. We refer the reader to Appendix C of \cite{Gaberdiel:2010jf} for more details on the Schottky parametrization.

The partition function of a genus-$g$ Riemann surface uniformised by the Schottky group is given by the following power series expansion in $p_i$ \cite{Gaberdiel:2010jf}:\footnote{Actually, the formula \eqref{Zg} just gives the partition function up to a factor of the form $e^{-cF_0}$; in other words, in an expansion of the free energy $F:=-\ln Z_g$ in $1/c$, it only gives the order $c^0$ and higher terms. The order-$c^1$ term $cF_0$ depends on the full metric on the Riemann surface, not just the complex structure. Its calculation within the context of the sewing construction is explained in appendix \ref{orderc}.}
\be\label{Zg}
Z_g=\sum_{h_1,\cdots,\,h_g}p_1^{h_1}\cdots p_g^{h_g}\;C_{h_1,\ldots,\,h_g}(a_3,\ldots,a_g,r_2,\ldots,r_g).
\ee
In the sewing construction, a handle is replaced by the boundary states inserted at the centers of the two disks. The functions $C_{h_1,\cdots,h_g}$ are $2g$-point functions on the Riemann sphere and $h_i$ is the conformal dimension of the operators inserted at the $i^{\rm{th}}$ pair of disks. A schematic picture of the sewing construction is shown in figure \ref{fig-sewing}. In the above equation we have fixed the positions $a_1=0$, $r_1=\infty$, and $a_2=1$. The $2g$-point functions $C_{h_1,\cdots,h_g}$, whose ingredients we will explain in the next paragraphs, are sums over products of vertex operators of the form
\be\label{Ch1hg}
C_{h_1,\cdots,h_g}=\sum_{\phi_i,\psi_i\in{{\cal H}_{h_i}}}\prod_{i=1}^{g}G_{\phi_i\psi_i}^{-1}\bigg\langle\prod_{i=1}^{g}V^{out}(\psi_i,r_i)\;\;V^{in}(\phi_i,a_i)\bigg\rangle,
\ee
where $G$ is the Zamolodchikov metric defined below in (\ref{G-ii}), and ${\cal H}_{h_i}$ is the Hilbert space of states of dimension $h_i$.

These expressions for the vertex operators should be understood as follows. Under any M\"obius transformation $\gamma(z)$, the vertex operator $V(\phi(z))$ transforms as \cite{Gaberdiel:1999mc}
\be\label{U}
V\bigg(U\Big(\gamma(z)\Big)\phi,\gamma(z)\bigg)=V\bigg(\gamma^{\prime}(z)^{L_0}\,e^{\frac{\gamma^{\prime\prime}(z)}{2\gamma^\prime(z)}\,L_1}\phi,\gamma(z)\bigg),
\ee
where
\be
\gamma^\prime(z)=\frac{d\gamma}{dz},\qquad\gamma^{\prime\prime}(z)=\frac{d^2\gamma}{dz^2}.
\ee
For the M\"obius transformation (\ref{mobius-i}) we have
\bea
\gamma_{a_i,r_i}^\prime(z)\!\!\!&=&\!\!\!\frac{(r_i-a_i)}{(z+1)^2},\qquad\gamma_{a_i,r_i}^{\prime\prime}(z)=-2\,\frac{(r_i-a_i)}{(z+1)^3},\label{gammaa}\nonumber\\
\Big(\gamma_{a_i,r_i}\hat\gamma\Big)^\prime(z)\!\!\!&=&\!\!\!\frac{(a_i-r_i)}{(z+1)^2},\qquad\Big(\gamma_{a_i,r_i}\hat\gamma\Big)^{\prime\prime}(z)=-2\,\frac{(a_i-r_i)}{(z+1)^3}.\label{gammar}
\eea
The ``in" and ``out" vertex operators in the sewing construction (\ref{Ch1hg}) then transform as
\be\label{Vin}
V^{in}(\phi_i,a_i)=V\bigg(U\Big(\gamma_{a_i,r_i}(z=0)\Big)\phi_i,\gamma_{a_i,r_i}(z=0)\bigg)=(r_i-a_i)^{L_0}e^{-L_1}\phi_i(a_i),
\ee
and 
\be\label{Vout}
V^{out}(\psi_i,r_i)=V\bigg(U\Big(\gamma_{a_i,r_i}\hat\gamma(z=0)\Big)\psi_i,\gamma_{a_i,r_i}\hat\gamma(z=0)\bigg)=(-1)^{L_0}(r_i-a_i)^{L_0}e^{-L_1}\psi_i(r_i).
\ee
For $i=1$, \emph{i.e.} for the handle with the two ends at $(0,\infty)$, one can perform an extra M\"obius transformation under which the two maps at zero and infinity become the identity and the inverse map, respectively. This is described in the next subsection. We note that if $\phi_i$ and $\psi_i$ are quasi-primaries, then the vertex operators are given by
\es{Vinout-qp}{
V^{in}(\phi_{qp},a_i)&=(r_i-a_i)^{h_{\phi_{qp}}}\phi_{qp}(a_i),\\
V^{out}(\psi_{qp},r_i)&=(-1)^{h_{\psi_{qp}}}(r_i-a_i)^{h_{\psi_{qp}}}\psi_{qp}(r_i).}

\subsubsection{Genus two}

We now specialize to genus-two Riemann surfaces. The partition function is given by
\be\label{Z2}
Z_{g=2}=\sum_{h_1,h_2=0}^{\infty}p_1^{h_1}p_2^{h_2}C_{h_1,h_2}(x),
\ee
where we have defined $r_2=x$. Using (\ref{Ch1hg}), the functions $C_{h_1,h_2}(x)$ are found to be
\eq{Ch1h2}{C_{h_1,h_2}(x)=\sum_{\phi_i,\psi_i\in{{\cal H}_{h_i}}}G_{\phi_1\psi_1}^{-1}G_{\phi_2\psi_2}^{-1}\bigg\langle V^{out}(\psi_1,\infty)\;V^{out}(\psi_2,x)\;V^{in}(\phi_2,1)\;V^{in}(\phi_1,0)\bigg\rangle~.}
These two formulae apply in general. For our purposes of computing $Z_{\rm vac}$, we only allow Virasoro descendants of the identity to be inserted at the boundary circles of the handles as in figure \ref{fig-sewing-vac}. Henceforth, we refer to \eqr{Z2} with the understanding that we compute $Z_{\rm vac}$ specifically. 

Let us now define the vertex operators needed in \eqr{Ch1h2}, starting with those at $(0,\infty)$. The functions $\Chh$ are invariant under the map $\gamma_{a_i,r_i}\to\gamma_{a_i,r_i}\gamma_{t}$, where $\gamma_t(z)=tz$, $t\in\bb C^*$. For the $i=1$ handle with its two ends located at $a_1=0$ and $r_1=\infty$, we consider a M\"obius transformation of the form $\gamma_{a_1,r_1}\gamma_{1/r_1}$ and find
\be\label{U0infty}
\gamma_{a_1,r_1}\gamma_{\frac1{r_1}}=\frac z{1+\frac z{r_1}}\Big|_{r_1\to\infty}=z,\qquad\qquad
\gamma_{a_1,r_1}\gamma_{\frac1{r_1}}\hat\gamma=\frac1{z+\frac1{r_1}}\Big|_{r_1\to\infty}=\frac1z.
\ee
This therefore gives the identity map for $a_1=0$ and the inverse map for $r_1=\infty$. The vertex operator at the origin is simply
\eq{V0}{V^{in}(\phi_1,0) = V(\phi_1,0) = \phi_1(0).}
The vertex operator at infinity follows from using
\be
\hat\gamma^\prime(z)=-\frac1{z^2},\qquad\hat\gamma^{\prime\prime}(z)=\frac2{z^3},
\ee
which yields
\be\label{U1/z}
V^{out}(\psi_1,\infty)=V\bigg(U\Big(\hat\gamma(z)\Big)\psi_1,\infty\bigg)=\lim_{z\to\infty}(-1)^{L_0}z^{2L_0}\,e^{-z\,L_1}\psi_1(z).
\ee
For the handle with vertices at $(a_2,r_2)=(1,x)$, we use (\ref{Vin})--(\ref{Vout}) to read off
\bea
V^{in}(\phi_2,1)\!\!\!&=&\!\!\!(x-1)^{L_0}e^{-L_1}\phi_2(1),\label{U1x-i}\\
V^{out}(\psi_2,x)\!\!\!&=&\!\!\!(-1)^{L_0}(x-1)^{L_0}e^{-L_1}\psi_2(x).\label{U1x-ii}
\eea

The Zamolodchikov metric is defined in terms of the in and out vertex operators as\footnote{We note that our formulae (\ref{Vout}) and (\ref{U1/z}) contain an extra factor of $(-1)^{L_0}$ comparing to the formulae in Appendix C of \cite{Gaberdiel:2010jf}. The reason is that we choose a different convention than that of \cite{Gaberdiel:2010jf}. In our convention $G$ is the Zamolodchikov metric whereas in \cite{Gaberdiel:2010jf} their metric $\hat G$ is a metric on the space of states which is related to $G$ via $\hat G_{\phi\psi}=G_{(-1)^{L_0}\phi\psi}$.}
\be\label{G-ii}
G_{\phi\psi}=\bigg\langle V^{out}(\psi,\infty) V^{in}(\phi,0)\bigg\rangle.
\ee
We choose an orthogonal basis of states at each level by diagonalizing the Gram matrix. The Zamolodchikov metric is thus diagonal and the ingoing and outgoing vertex operators are the same up to M\"obius transformations. Consequently, we can (and will) define the norm of the states as $\cl N_{\phi}\equiv G_{\phi\psi}$. 

To summarize, the following is the prescription for constructing the genus-two partition function: insert vertex operators \eqr{V0}, (\ref{U1/z})--(\ref{U1x-ii}) and the Zamolodchikov metric \eqr{G-ii} into (\ref{Ch1h2}) to evaluate $C_{h_1,h_2}(x)$, and sum over these using (\ref{Z2}).
\subsection{Four-point functions $C_{h_1,h_2}(x)$}\label{iv-ii}
We will momentarily compute some of the functions $C_{h_1,h_2}(x)$ defined in (\ref{Ch1h2}). Before doing so, it is useful to elucidate some of their general properties.

First, these functions are symmetric under the exchange of the positions of the two handles:
\be
C_{h_1,h_2}(x)=C_{h_2,h_1}(x)
\ee
When $h_1=0$ or $h_2=0$,  
\be\label{c0h}
C_{0,h}(x)  = C_{h,0}(x) = d(h)~,
\ee
where $d(h)$ is the degeneracy of operators at level $h$. This follows from the definition of the vertex operators and of the $C_{h_1,h_2}(x)$ themselves. It can also be understood intuitively: replacing either handle with two insertions of the identity reduces the genus-two partition function to the torus partition function, the holomorphic half of which is $\Tr p^{L_0} = \sum_h d(h) p^h$.\footnote{Note the $c$-independence of this quantity. When summing over the vacuum module only, the dimensions of each state are fixed by conformal symmetry, and hence unrenormalized. This is the CFT statement of the one-loop exactness of the pure gravity partition function on a solid torus.}
Thus, \eqr{Z2} implies (\ref{c0h}). 

Now we consider the $x$-dependence of $\Chh$. For general $x$, they obey 
\be
C_{h_1,h_2}(x) = C_{h_1,h_2}(1/x)
\ee
which is a consequence of modularity with respect to $Sp(4,\mathbb{Z})$. Taking various limits in $x$ corresponds to taking OPE limits of the (dressed) four-point functions defining $C_{h_1,h_2}(x)$. The simplest one is $x\rightarrow1$, which describes the fusion of two ends of the same handle. In this case, 
\be\label{cx1}
C_{h_1,h_2}(1) = d(h_1)\times d(h_2)~.
\ee
This again follows by definition, and is necessary for the partition function \eqr{Z2} to factorize in the separating degeneration limit.

More subtle are the equivalent OPE limits $x\rightarrow 0$ and $x\rightarrow\infty$, which describes the fusion of two ends of different handles. These limits are singular. In Appendix \ref{app-OPE}, we show that in a $1/c$ expansion, the leading powers of $x\rar 0$ that appear are correlated with powers of $1/c$ as follows:
\be\label{pc2}
\lim_{c\rightarrow\infty}\lim_{x\rightarrow0}C_{h_1,h-h_1}(x) \sim O(x^{-h})+ {1\over c} O(x^{-h+2}) + \left(\sum_{n=2}^{\infty}{1\over c^n}\right)O(x^{-h+4}).
\ee
We have defined $h_2=h-h_1$, and are assuming $h>h_1>0$ because $C_{0,h}(x)$ is constant. We are ignoring $h_1$- and $h$-dependent coefficients at each order, and displaying only the leading singular behavior at each order in $1/c$. The last term means that at $\cO(1/c^2)$ and {\it all} orders beyond, the leading divergence scales as $O(x^{-h+4})$.

We now proceed to compute $C_{h_1,h_2}(x)$ explicitly for low values of $h_1$ and $h_2$. A word on notation: henceforth, we denote the set of operators at level $h$ above the ground state as $\lbrace \Oc_h^{(i)}\rbrace$, where $i=1,2,\ldots, d(h)$. 

\subsubsection{$C_{h,0}(x)$}\label{iv-ii-i}

In this case, the identity operator propagates through one of the handles, so the four-point functions reduce to two-point functions
\es{C0h}{
C_{h,0}(x)&=\sum_{\phi,\psi\in{{\cal H}_{h}}}G_{\phi\psi}^{-1}\bigg\langle V^{out}(\psi,\infty)\;V^{in}(\phi,0)\bigg\rangle,\\
C_{0,h}(x)&=\sum_{\phi,\psi\in{{\cal H}_{h}}}G_{\phi\psi}^{-1}\bigg\langle V^{out}(\psi,x)\;V^{in}(\phi,1)\bigg\rangle.}
As discussed earlier, $C_{h,0}(x)=d(h)$; from the definition of $G_{\phi\phi}$ in \eqr{G-ii}, this is obviously true. It is less obvious that $C_{0,h}(x)=d(h)$ by looking at \eqr{C0h}. Hence, we find it instructive to compute two $C_{0,h}(x)$ in detail, to illustrate how to use the method of \cite{Gaberdiel:2010jf} outlined above. In the first example we consider a quasi-primary state and in the second example we consider a secondary state. The latter is particularly useful, as secondary operators transform nontrivially under the M\"obius transformations, so care must be taken in computing their correlation functions. 

At level 2 ($h=2$) there is only one state, the stress tensor $T=L_{-2}|0\rangle$, which is a quasi-primary with norm ${\cal N}_T=\frac c2$. From \eqr{U1x-i}--\eqr{U1x-ii}, the vertex operators at $x$ and $1$ are
\es{Tvertex}{V^{out}(T,x) &= (x-1)^2 \,T(x),\\
V^{in}(T,1) &= (x-1)^2 \,T(1).\\
}
Then we have, as expected,
\es{C02}{
C_{0,2}(x)&= {\cal N}_T^{-1}\bigg\langle V^{out}(T,x)\;V^{in}(T,1)\bigg\rangle  = 1~,}
where we used the stress tensor two-point function 
\eq{TT}{\langle T(x) T(1) \rangle = {c\over 2}{1\over (x-1)^4}~.}

At level 3, there is again one state $\cO_3=\p T=L_{-3}|0\rangle$, this time a secondary state with norm ${\cal N}_{\cO_3}=2c$. The vertex operators at $x$ and $1$ are now, from \eqr{U1x-i}--\eqr{U1x-ii},
\es{}{V^{out}(\Oc_3,x) &= -(x-1)^3\p T(x)-4(x-1)^2T(x),\\
V^{in}(\Oc_3,1) &= (x-1)^3\p T(1)-4(x-1)^2T(1).}
Using \eqr{TT} we again have, as expected,
\bea\label{C03}
C_{0,3}(x)\!\!\!&=&\!\!\!\cl N_{\cO_3}^{-1}\bigg\langle V^{out}(\cO_3,x)\;V^{in}(\cO_3,1)\bigg\rangle\nonumber\\
\!\!\!&=&\!\!\!\frac1{2c}\bigg\langle \Big(-(x-1)^3\,\partial T(x)-4(x-1)^2\,T(x)\Big)\;\Big((x-1)^3\partial\,T(1)-4\,(x-1)^2T(1)\Big)\bigg\rangle\nonumber\\
\!\!\!&=&\!\!\!1.
\eea
It is also useful to see the vertex operator at infinity, which carries non-trivial dressing on account of $\Oc_3$ being a secondary operator:
\eq{L-3i}{V^{out}(\Oc_3,\infty) = \lim_{z\to\infty} V\Big((-1)^{L_0}z^{2L_0}\,e^{-z\,L_1}\Oc_3,z\Big) = \lim_{z\rar\infty} \Big(-z^6 \p T(z) - 4z^5 T (z)\Big).}

We next move on to the computation of $C_{h_1,h_2}(x)$ for $h_1\ne0$ and $h_2\ne0$. We compute the requisite four-point functions using the methods described in section \ref{iii}. The transformation properties of the vertex operators are evaluated following the same procedure shown in the above examples; accordingly, the presentation here is streamlined, with some details of the vertex operator transformations relegated to Appendix \ref{app-Cs}. There, we also list the operators and their norms through level six of the vacuum module. 

\subsubsection{$C_{h,2}(x)$}\label{iv-ii-i}

First, consider the four-point function with $h_2=2$ which corresponds to the four-point function of four stress tensors, each dressed with appropriate factors of $x$ and $z$ (recall equation \ref{Ch1h2}). We find that
\bea\label{C22}
C_{2,2}(x)\!\!\!&=&\!\!\!\frac1{\left(\frac c2\right)^2}\lim_{z\to\infty}\bigg\langle z^{4}T(z)\;\;(x-1)^2\,T(x)\;\;(x-1)^2T(1)\;\;T(0)\bigg\rangle\nonumber\\
\!\!\!&=&\left(1+(x-1)^4+\frac{(x-1)^4}{x^4}\right)+\frac8c\,\frac{(x-1)^2\,(1-x+x^2)}{x^2},
\eea
We note that $C_{2,2}(x)$ is manifestly symmetric under $x\to1/x$, as required.

We next compute $C_{3,2}(x)$. This can be done by taking the derivative of $C_{2,2}(x)$ with respect to the sphere coordinates at the insertion points of $\cO_3$, or by direct computation of the four-point function using the Virasoro mode expansion formula \eqr{eq8}. We obtain
\bea\label{C32}
C_{3,2}(x)\!\!\!&=&\!\!\!\frac1{x^5}\,\Big(4-16x+24x^2-16x^3+4x^4+x^5+4x^6-16x^7+24x^8-16x^9+4x^{10}\Big)\nonumber\\
\!\!\!&+&\!\!\!\frac1c\,\frac{2\,(x-1)^2}{x^3}\,\Big(4+x-4x^2+x^3+4x^4\Big),
\eea
We used \eqr{Tvertex} and (\ref{L-3i}) to write down the necessary vertex operators. One can explicitly check that $C_{3,2}(x)=C_{2,3}(x)$, as required.

We next move to level four and evaluate $C_{4,2}(x)$. There are two orthogonal states at level four,
\be\label{O4}
\cO_{4}^{(1)}=\L = \left(L_{-2}^2-\frac35L_{-4}\right)|0\rangle~, \quad \cO_4^{(2)}=L_{-4}|0\rangle~,
\ee
where $\L$ is the quasi-primary first recalled in \eqr{Lamb}, and $\cO_4^{(2)}=\p^2 T/2$ is secondary with norm ${\cal N}_{\cO_4^{(2)}}=5c$. We obtain 
\bea\label{C42}
C_{4,2}(x)\!\!\!&=&\!\!\!\frac2{x^6}\,\Big(5 - 20 x + 31 x^2 - 24 x^3 + 11 x^4 - 4 x^5 + 3 x^6\\
&&\quad - 4 x^7+11 x^8-24 x^9 + 31 x^{10}-20 x^{11} + 5 x^{12}\Big)\nonumber\\
&&\!\!\!\!\!\!\!\!\!\!\!+\frac1c\,\frac{4\,(x-1)^2}{x^4}\,\Big(4 - 3 x + 10 x^2 - 14 x^3 + 10 x^4 - 3 x^5 + 4 x^6\Big).\nonumber
\eea
Note that the finite expansion in $1/c$ is at first surprising because ${\cal N}_{\L}^{-1}$ has an infinite $1/c$ expansion; there are non-trivial cancellations between the two correlators. We will explain this in a moment.

For the remaining computations, we will be briefer. (We remind the reader of Appendix \ref{app-Cs} containing further details.) At level five, we find
\bea\label{C25}
C_{5,2}(x)\!\!\!&=&\!\!\!\frac1{x^7}\,\Big(20-80x+124x^2-95x^3+40x^4-10x^5\\
&&\quad+4x^7-10x^9+40x^{10}-95x^{11}+124x^{12}-80x^{13}+20x^{14}\Big)\nonumber\\
\!\!\!&+&\!\!\!\frac1c\,\frac{4\,(x-1)^2}{x^5}\,\Big(6-6x+7x^2+5x^3-14x^4+5x^5+7x^6-6x^7+6x^8\Big).\nonumber
\eea
At level six, we find
\bea\label{C26}
&&\!\!\!\!\!\!\!\!\!\!C_{6,2}(x)=\frac1{x^8}\,\Big(35-140x+220x^2-172x^3+67x^4-8x^5+2x^6-8x^7+12x^8\\
&&\qquad\qquad\,-8x^9+2x^{10}-8x^{11}+67x^{12}-172x^{13}+220x^{14}-140x^{15}+35x^{16}\Big)\nonumber\\
&&\qquad+\frac1c\,\frac{4\,(x-1)^2}{x^6}\,\Big(8-7x+10x^2-8x^3+46x^4-74x^5+46x^6-8x^7\nonumber\\
&&\qquad\qquad\qquad\qquad\;+10x^8-7x^9+8x^{10}\Big).\nonumber
\eea

We observe that all the functions $C_{h,2}(x)$ we have computed so far contain a term proportional to $1/c$ and a term constant in $c$. In fact, this is true for all $h$. The reason is that the sum over $h$ of $C_{h,2}(x)$ is related to the two-point function of the stress tensor on the torus with the insertion of a vacuum projector: 
\begin{equation}
\ev{P_{\rm vac}T(1)T(x)} = \frac c2\frac1{(x-1)^4}\sum_hC_{h,2}(x)p_1^h
\end{equation}
(where $p_1=e^{2\pi i \tau}$). This two-point function can in turn be obtained by differentiating the vacuum free energy $F_{\rm vac}(T^2)$ with respect to the metric. This free energy was shown at the end of subsection \ref{powers} to be one-loop exact, in other words to contain only terms linear and constant in $c$. Hence, for all $h$, $C_{h,2}(x)$ must contain only terms constant and inversely proportional to $c$.

\subsubsection{$C_{h,3}(x)$}\label{iv-ii-i}

We next compute the functions $C_{h_1,h_2}(x)$ with $h_1\ge3$ and $h_2=3$. The first function is $C_{3,3}(x)$ which corresponds to the four-point function of four $\partial T$'s. This correlation function can be evaluated by taking the derivatives of $C_{2,2}(x)$ with respect to the sphere coordinates at the location of the four operators, or by direct computation. We find
\bea\label{C33}
C_{3,3}(x)\!\!\!&=&\!\!\!\frac{1}{x^6}\,\Big(25-110x+191x^2-164x^3+71x^4-14x^5\\
&&\quad+3x^6-14x^7+71x^8-164x^9+191x^{10}-110 x^{11}+25x^{12}\Big)\nonumber\\
&&\!\!\!\!\!\!\!\!\!\!+\frac1c\,\frac{(x-1)^2}{x^4}\,\Big(18-6x+x^2-8x^3+x^4-6x^5+18x^6\Big),\nonumber
\eea
Next, for $C_{4,3}(x)$ we find
\bea\label{C34}
&&\!\!\!\!\!\!\!\!\!\!C_{4,3}(x)=\frac2{x^7}\,\Big(45-210x+399x^2-396x^3+219x^4-66x^5+9x^6\\
&&\qquad\qquad+x^7+9x^8-66x^9+219x^{10}-396x^{11}+399x^{12}-210x^{13}+45x^{14}\Big)\nonumber\\
&&\qquad+\frac1c\,\frac{(x-1)^2}{x^5}\,\Big(64-83x+68x^2+3x^3-56x^4+3x^5+68x^6-83x^7+64x^8\Big),\nonumber
\eea
and for $C_{5,3}(x)$ we find
\bea\label{C35}
&&\!\!\!\!\!\!\!\!\!\!C_{5,3}(x)=\frac1{x^8}\,\Big(245-1190x+2380x^2-2526x^3+1530x^4\\
&&\qquad\qquad-530x^5+100x^6-10x^7+4x^8-10x^9+100x^{10}\nonumber\\
&&\qquad\qquad-530x^{11}+1530x^{12}-2526x^{13}+2380x^{14}-1190x^{15}+245x^{16}\Big)\nonumber\\
\!\!\!&+&\!\!\!\frac1c\,\frac{(x-1)^2}{x^6}\,\Big(150-264x+201x^2-32x^3+3x^4-56x^5+3x^6\nonumber\\
&&\qquad\qquad\;-32x^7+201x^8-264x^9+150x^{10}\Big).\nonumber
\eea

Again, all of the functions $C_{h,3}(x)$ evaluated so far truncate at order $1/c$ in a large-$c$ expansion. This is because all four-point functions in $C_{h,3}(x)$ can be computed by taking derivatives of stress tensors in $C_{h,2}(x)$, which do not affect the $c$-dependence of the correlators.

\subsubsection{$C_{4,4}(x)$}\label{iv-ii-i}

The function $C_{4,4}(x)$ is a linear combination of four four-point functions:
\be\label{C44}
C_{4,4}(x)=\sum_{i,j=1}^2\N_{\cO_{4,i}}^{-1}\N_{\cO_{4,j}}^{-1}\bigg\langle V^{out}(\cO_{4,i},\infty)\;V^{out}(\cO_{4,j},x)\;V^{in}(\cO_{4,j},1)\;V^{in}(\cO_{4,i},0)\bigg\rangle~.
\ee
By the same argument below \eqr{C25} and \eqr{C35}, the terms which contain at least one pair of the secondary operator $\cO_4^{(2)}=\p^2T/2$ truncate at order $1/c$ in a large-$c$ expansion. The only term in $C_{4,4}(x)$ which could potentially contribute at higher orders in $1/c$ is the four-point function of four quasi-primaries $\L$, defined in \eqr{Lamb}. Let us focus on this contribution, which we call $C_{4,4}|_{\L}(x)$.

Using the definitions of the vertex operators simply yields
\eq{446}{C_{4,4}|_{\L}(x) = \N_{\L}^{-2}(x-1)^8\lim_{z\rar\infty} z^8\bigg \langle \L(z)\, \L(x)\, \L(1) \,\L(0) \bigg\rangle~.}
The norm of $\L$ was given in \eqr{No4}. Substituting this and the results \eqr{F4h}--\eqr{348} for the four-point function obtained via the holomorphic bootstrap, we find
\bea\label{C44-qp}
C_{4,4}|_{\L}(x)\!\!\!&=&\!\!\!\frac{(1-x+x^2)^8}{x^8}+\left(\frac{32}c-8\right)\frac{(x-1)^2(1-x+x^2)^5}{x^6}\nonumber\\
\!\!\!&+&\!\!\!\frac{4\,(3704+590c+125c^2)}{5c\,(22+5c)}\frac{(x-1)^4(1-x+x^2)^2}{x^4}~.
\eea
Crucially to what follows, we observe that $C_{4,4}|_{\L}(x)$ contributes to an infinite expansion in $1/c$. This comes entirely from the inverse norms in \eqr{446}. 
\vs

The collection of $\Chh$ computed in this subsection, plugged into the partition function \eqr{Z2}, forms one of our main computational results: namely, the first several terms in the Virasoro vacuum module contribution to the partition function of an arbitrary CFT on a genus-two Riemann surface, in the regime of Schottky parameters $p_1,p_2\ll1$. 

\section{Free energy at large central charge, 3D gravity, and R\'enyi entropies}\label{v}

Having derived the first handful of terms in equation \eqr{Z2} for all $c$, it is trivial to expand it at large $c$. As we have discussed, this large $c$ expansion may be interpreted as the semiclassical expansion of the pure 3D quantum gravity partition function around genus-two handlebody geometries with conformal boundary $\S$ specified by Schottky parameters $\lbrace p_1,p_2,x\rbrace$. In subsection \ref{v-i}, we present some of our main results. First, we provide explicit contributions of the Virasoro vacuum representation to the CFT free energy at all orders in $1/c$, corresponding to all-loop free energies in the gravitational loop expansion. We also show that at least in the perturbative regime $p_1,p_2\ll1$, the loop expansion does not truncate except when $\S$ is the union of two tori.

We then proceed to subsections \ref{v-ii} and \ref{v-iii}, where we expand our general result \eqr{Z2} near two symmetric points in the genus-two moduli space: the replica surface $\R_{2,3}$ used to compute the R\'enyi entropy $S_3$ for two disjoint intervals in vacuum, and the point corresponding to the separating degeneration limit $x=1$. Our results will extend those of \cite{Chen:2013dxa} and \cite{Yin:2007gv}, respectively.

\subsec{All-loop results in 3D quantum gravity}\label{v-i}
We consider the $1/c$ expansion of the vacuum free energy, $F_{\vac}=-\log Z_{\vac}$:
\eq{}{F_{\vac} = \sum_{\ell=0}^{\infty} c^{1-\ell}F_{\vac;\,\ell}}
where $\ell$ denotes the loop order. Its moduli-dependence is kept implicit. We can read off the loop corrections $F_{\vac ;\,\ell}$ from \eqr{Z2} upon expanding the $\Chh$ in $1/c$. We likewise expand these as
\eq{chhell}{\Chh = \sum_{\ell=1}^{\infty}c^{1-\ell}C_{h_1,h_2;\,\ell}(x)~.}

To begin, note that in the small $(p_1,p_2)$ expansion in which we work, both the one- and two-loop free energies are nonzero. This follows from the explicit results in section \ref{iv} and from the R\'enyi entropy computation \eqr{chenpatt}, but also from our general exposition of $c$-scaling of identity module correlators in section \ref{powers}. 

More interesting is the question of whether there are higher-loop terms. For $\ell>2$, the $\Chh$ that we have computed all obey $C_{h_1,h_2;\,\ell}(x)=0$ except for $C_{4,4}|_{\L}(x)$, computed in \eqr{C44-qp}, which clearly has an infinite expansion. Accordingly, the leading contribution to the three-loop free energy $F_{\vac;\,3}$ in a small $(p_1,p_2)$ expansion can, and does, appear at $O(p_1^4p_2^4)$:
\eq{f30}{F_{\vac;\,3}= p_1^4p_2^4\left(C_{4,4;\,3}(x) - \half \left(C_{2,2;\,2}(x)\right)^2\right) + O(p_1^4p_2^5)~.}
There is no cancellation: instead, our results yield
\eq{f3}{F_{\vac;\,3} = p_1^4p_2^4\,{13312 (x-1)^4(1-x+x^2)^2\over 25x^4} + O(p_1^4p_2^5)~.}
For $x\in \mathbb{R}$, this only vanishes at $x=1$. This point in moduli space corresponds to the strict separating degeneration limit. As the torus free energy is known to truncate at $\Oc(c^0)$ in a large $c$ expansion \cite{Maloney:2007ud}, the fact that $F_{\vac;\,3}=0$ when $x=1$ is required by consistency. The interesting result proven here is that $F_{\vac;\,3}$ is nonzero everywhere else on the real line.\footnote{Note that \eqr{f30} and \eqr{f3} also hold when $p_1=p_2$, because the only contribution at $\Oc(1/c^2)$ to $\Chh$ for $h_1+h_2=8$ comes from $C_{4,4}(x)$. On the other hand, if $x$ is a function of $p_1$ and $p_2$, higher order terms in \eqr{f3} are not necessarily suppressed. We will encounter such a situation in our discussion of R\'enyi entropy.} 

In fact, the infinite $1/c$ expansion of $C_{4,4}(x)$ and the finite $1/c$ expansion of $C_{2,2}(x)$ together imply that the $\ell$-loop free energy $F_{\vac;\,\ell}$ is nonzero for {\it all} $\ell$, at least in a small $(p_1,p_2)$ expansion. The reason is simply that the term at $O(p_1^4p_2^4)$ cannot be cancelled by higher order terms in $p_1$ and $p_2$. For $\ell>3$---that is, at $\Oc(1/c^3)$ and beyond---$F_{\vac;\,\ell}$ is given to leading order in $p_1,p_2$ as
\eq{allloop}{F_{\vac;\,\ell>3} = p_1^4p_2^4 ~C_{4,4;\,\ell>3}(x) + O(p_1^4p_2^5)}
with 
\eq{c44e}{C_{4,4;\,\ell>3}(x) = { (x-1)^4(1-x+x^2)^2\over x^4}\cdot \left({4(3704+590c+125c^2)\over5c(5c+22)}\right)\Bigg|_{c^{1-\ell}}~.}
determined by \eqr{C44-qp}. From a technical standpoint, this non-trivial $1/c$ expansion arises from the inverse norms appearing in the sewing construction, cancelling the norms in the correlator $\langle \L\L\L\L\rangle$. 

Interpreted as a CFT result, \eqr{allloop} is an exact expression for the contribution of the Virasoro vacuum module to the genus-two free energy of any family of CFTs that admits a $1/c$ expansion. Interpreted as a pure gravity result, \eqr{allloop} is an explicit formula for all-loop free energies on genus-two handlebodies. The loop counting parameter in the bulk is $G_N=3R_{\rm AdS}/2c$. In contrast to the one-loop exactness at genus one, the genus-two free energy is not exact at {\it any} loop order.

Strictly speaking, we have so far established that the semiclassical expansion does not truncate for any real $x\neq 1$. What about complex $x$? In particular, the all-loop terms \eqr{f3} and \eqr{c44e} clearly vanish at $x=e^{\pm i\pi/3}$, the complex roots of $1-x+x^2=0$. This follows from the same property of the four-point function of $\L$, as discussed below equation \eqr{348}. But this is a special feature of correlators of identical operators with $h/3\notin \mathbb{Z}$, so it will not persist to higher orders in the sewing expansion. For instance, $C_{4,5;\,\ell>2}(e^{\pm i\pi/3})\neq 0$, and likewise at all higher levels. Therefore, we have shown the following statement: {\it perturbatively in $(p_1,p_2)$, the loop expansion does not truncate for any genus-two handlebodies except at the separating degeneration point.}

We note that at fixed order in $p_1$ and $p_2$, the $1/c$ expansion for $Z_{\rm vac}$ converges. This does not necessarily imply, however, that at a fixed point in moduli space (i.e.\ for fixed values of $p_1$ and $p_2$) the $1/c$ expansion converges.  Indeed, since there are presumably other saddle point contributions to the path integral (coming from bulk handlebodies with different topology) one might expect that the series expansion of $Z_{\rm vac}$ is asymptotic in $1/c$.  However, in the genus one case the series expansion for $Z_{\rm vac}$  converges---in fact, it truncates at order $c^0$.   Thus it is an interesting open question whether the $1/c$ expansion for $Z_{\rm vac}$ converges at higher genus.

Finally, let us comment on the holographic interpretation of our result for the one-loop partition function, $Z_{\vac;\,1}$. This can be viewed as a computation of the holomorphic half of the graviton handlebody determinant,\footnote{The first product in \eqr{grav} runs over primitive elements $\gamma\in{\cal P}\subset\Gamma$, defined as those elements that cannot be written as $\gamma=\b^m$ for $\b \in \Gamma$ and $m>1$. The eigenvalues of $\gamma$ are ${\rm eig}(\gamma) = q_{\gamma}^{\pm 1/2}$, and we do not count $\gamma$ and $\gamma^{-1}$ as distinct elements.} 
\eq{grav}{Z_1^{\rm grav} = \prod_{\gamma \in {\cal P}} \prod_{n=2}^{\infty} {1\over |1-q_{\gamma}^n|^2}~,}
The novelty of our computation is that we work in the regime of $p_1,p_2\ll 1$ but for arbitrary $x$. This regime has not been probed directly in existing computations of \eqr{grav}. In \cite{Barrella:2013wja}, \eqr{grav} was computed for handlebodies asymptotic to replica manifold for two-interval R\'enyi entropy in a short interval expansion; this has only a single modulus and requires $p_1=p_2 \ll 1$ and $x \gg 1$. In \cite{Yin:2007gv}, \eqr{grav} was computed near the separating degeneration limit where $\S$ becomes the union of two tori, which requires $x \approx 1$ for arbitrary $(p_1,p_2)$.  

\subsubsection{Higher spin theories}\label{v-i-i}
So far, we have restricted to the pure Virasoro sector of the CFT. The meaning and calculation of $Z_{\vac}$ are conceptually unmodified in the presence of higher spin currents. Along with the stress tensor, these Virasoro primaries live in the vacuum representation of an extended conformal symmetry, typically a $W$ algebra. In the computation of $Z_{\vac}$ by sewing, we now allow these currents and their normal ordered products to propagate through the handles. The resulting $Z_{\vac}$ is again of the form \eqr{Z2}, only with different coefficients $\Chh$. 

The holographic dual of $Z_{\vac}$ in the presence of higher spin symmetry is the perturbative partition function of pure 3D higher spin gravity. A bulk Chern-Simons theory with connections valued in two copies of a Lie algebra ${\cal G}$ describes the vacuum sector of a CFT whose $W$ algebra is the Drinfeld-Sokolov reduction of ${\cal G}$ \cite{Campoleoni:2010zq}. Accordingly, the $1/c$ expansion of $Z_{\vac}$ for such a CFT yields the semiclassical loop expansion of the ${\cal G}\times{\cal G}$ Chern-Simons higher spin theory.

As a simple example, consider a CFT with $W_{3}$ symmetry, which contains a single higher spin current of spin three, $W$. Its presence will modify most of the $\Chh$ coefficients, starting with $C_{3,2}(x)$ and $C_{3,3}(x)$.\footnote{The generating function of quasi-primaries containing at least one $W$ current is given in Appendix B of \cite{Perlmutter:2013paa}.} We can easily compute these using the correlators of section \ref{iii}. The interesting term is $C_{3,3}(x)$. Denoting the contribution to $C_{3,3}(x)$ from the $W$ current four-point function as $\delta_W C_{3,3}(x)$, we find, using \eqr{F3} and \eqr{wcoeffs},
\es{}{\delta_WC_{3,3}(x) &:= \N_W^{-2} (x-1)^6 \lim_{z\rar\infty} z^6 \langle W(z) W(x) W(1) W(0)\rangle \\
&= {(1-x+x^2)^6\over x^6} + {6(3-c)\over c} {(1-x)^2  (1-x+x^2)^3\over x^4}\\& +  {3(5c^2-71c-102)\over c(5c+22)} {(1-x)^4 \over x^2}~.}
Expanded at large $c$, this yields an infinite series of loop corrections to the free energy of ${\cal G}=SL(3)$ higher spin gravity:
\eq{}{F_{\vac;\,\ell>2}^{\rm SL(3)} = p_1^3p_2^3 \,{(1-x)^4 \over x^2}\cdot \left({3(5c^2-71c-102)\over c(5c+22)}\right) \Bigg|_{c^{1-\ell}} + O(p_1^3p_2^4)~.}
This is nonzero for all $x\neq 1$, so we conclude that the loop expansion does not truncate away from the separating degeneration limit for small $(p_1,p_2)$. This is true for all higher spin algebras ${\cal G}$. We will return to the topic of higher spin theories in the Discussion.

\subsec{R\'enyi entropies}\label{v-ii}
As discussed in section \ref{ii}, there are three R\'enyi entropies that involve genus-two replica manifolds without punctures (i.e. for CFTs not in excited states). These are the $N=2,n=3$ and $N=3,n=2$ R\'enyi entropies for a CFT on the plane, and $N=1,n=2$ for a CFT on the torus. We mostly focus on the $N=2$ case, with replica manifold $\R_{2,3}$. Our results in section \ref{v-i} are sufficient to rule out the truncation of the $1/c$ expansion of $F_{\vac}$ even in the case of the replica manifold $\S=\R_{2,3}$ introduced in section \ref{Renyireview}. We now exhibit this in detail; the final results can be found in \eqr{f3ren} and \eqr{renall}.

\subsubsection{Two intervals on the plane}\label{v-iia}

Our goal is to express the free energy in terms of the coordinate $y$ parameterizing the interval spacing, defined in section \ref{ii}. To do so, we need only to express the Schottky coordinates $\lbrace p_1,p_2,x\rbrace$ in terms of $y$. One way to proceed is by using the period matrix $\Om(y)$ for the replica manifold $\R_{2,3}$, which is known \cite{Calabrese:2009ez}. Thus, we will perform the map $\lbrace p_1,p_2,x\rbrace \mapsto \lbrace q_{ij}(y)\rbrace$, where $q_{ij}(y) = \exp[2\pi i \Om_{ij}(y)]$ are the multiplicative periods, in the regime of $y$ corresponding to small $(p_1,p_2)$. Plugging into \eqr{Z2} gives $F_{\vac}(\R_{2,3})$ for arbitrary $c$; we then proceed to study this result at large $c$. 

For two disjoint intervals and arbitrary $n$, the period matrix is \cite{Calabrese:2009ez}
\es{rnprd}{
\Om_{ij}(y)=\frac{2i}{n}\sum_{k=1}^{n-1}\sin\left(\pi\frac{k}{n}\right)\,\cos\left(2\pi\frac{k}{n}(i-j)\right)\frac{_2F_1\left(\frac{k}n,1-\frac{k}n;1;1-y\right)}{_2F_1\left(\frac{k}n,1-\frac{k}n;1;y\right)}~,}
Specializing to $n=3$, the period matrix is given by
\es{rnprd3}{
\Om(y)=\frac{2i}{\sqrt3}\,\frac{_2F_1\left(\frac13,\frac23;1;1-y\right)}{_2F_1\left(\frac13,\frac23;1;y\right)}\left({\begin{array}{cc}
1 & -\frac12 \\
-\frac12 & 1 \\
\end{array} } \right).}
This is a highly symmetric genus-two Riemann surface: there is only a single modulus $y$, as opposed to the $3g-3=3$ moduli of a generic genus-two surface.

To express the Schottky coordinates in terms of $y$, we need to invert the power series expansion given in \eqr{mltprds}. The fact that $q_{11}=q_{22}$ implies that in Schottky coordinates, $\R_{2,3}$ has $p_1=p_2\equiv p$, as a quick inspection of \eqr{mltprds} reveals. Our results are applicable when $p\ll 1$, so \eqr{mltprds} forces us to take $q_{11}\ll 1$ too. From \eqr{rnprd3}, this is just the short interval limit, $y\ll 1$, often studied in the context of 2D CFT R\'enyi entropy: taking $y\ll1$ in \eqr{rnprd3} yields multiplicative periods
\es{qy}{&q_{11}\big|_{y\ll1}=\frac{y^2}{729}+\frac{10y^3}{6561}+\frac{29y^4}{19683}+O(y^5),\\
&q_{12}\big|_{y\ll1}=\frac{27}{y}-15-2y-\frac{734y^2}{729}-\frac{4181y^3}{6561}+O(y^4)~.}
Finally, we obtain the series expansion of $p$ and $x$ in terms of $y$ by inverting \eqr{mltprds} using \eqr{qy} and the explicit results for the coefficients $c(n,m,|r|)$ and $d(n,m,r)$ given in \cite{Gaberdiel:2010jf}. The result is
\bea\label{pxy-sub}
p(y)\!\!\!&=&\!\!\!\frac{y^2}{729}+\frac{28}{19683}\,y^3+\frac{26}{19683}\,y^4+\frac{5768}{4782969}\,y^5 +\frac{47429}{43046721}\,y^6+\frac{10582844}{10460353203}\,y^7+O(y^8),\nonumber\\
x(y)\!\!\!&=&\!\!\!\frac{27}{y}-15-\frac{56}{27}\,y-\frac{28}{27}\,y^2-\frac{12892}{19683}\,y^3-\frac{3044}{6561}\,y^4 +O(y^5)\,.
\eea
Note that $x(y)$ diverges linearly for small $y$.\footnote{We note that $p(y)$ is nothing but the square of the larger eigenvalue of the Schottky generators themselves: $\text{eig}(L_i(y)) = p(y)^{\pm 2}$, where $L_1(y) = L_2(y)$ are the Schottky generators in the $y\ll 1$ regime. This was already computed in \cite{Barrella:2013wja,Perlmutter:2013paa,Chen:2013dxa,Beccaria:2014lqa}; see in particular equation (3.8) of \cite{Beccaria:2014lqa}, with $k=1, n=3$. One can then find $x(y)$ using the Schottky relations. Such an algorithm is an alternative to that presented in the text.}

We can now compute the vacuum free energy $F_{\vac}(y)=-\log Z_{\vac}(y)$, and hence the vacuum contributions to R\'enyi entropy, in a $y\ll 1$ short interval expansion, where
\eq{zy}{Z_{\vac}(y) = \sum_{h_1,h_2=0}^{\infty} p(y)^{h_1+h_2}C_{h_1,h_2}(y)~,}
We will further expand this result at large $c$ and compare to those of \cite{Chen:2013dxa}. 

In order to perform this expansion, we need to be a bit careful: because powers of $x$ introduce inverse powers of $y$, it is not manifest in \eqr{zy} that the short interval expansion can be meaningfully organized in powers of $p(y)$. We need to know something about how $C_{h_1,h_2}(x)$ scales with large $x$, and hence small $y$. Fortunately, we can read this off from \eqr{pc2}. Keeping terms to leading order in $y\rar 0$ at each order in $1/c$, and ignoring coefficients, \eqr{pc2} and \eqr{pxy-sub} imply that for $h>h_1>0$, 
\eq{pc1}{\lim_{c\rar\infty}\lim_{y\rar 0}\, p(y)^hC_{h_1,h-h_1}(y) \sim O(y^{h})+ {1\over c} O(y^{h+2}) + \left(\sum_{n=2}^{\infty}{1\over c^n}\right)O(y^{h+4})~.}
Therefore, we can indeed ignore higher order terms in the sum over $h$ when we expand in small $y$. 

Without further ado, the results are as follows. At $\ell=1,2$ we find
\es{MI}{F_{\vac;\,1}(y)&=\frac{y^4}{177147}+\frac{56\,y^5}{4782969}+\frac{2189\,y^6}{129140163}+\frac{24668\,y^7}{1162261467}+O(y^8)\\
F_{\vac;\,2}(y)&=\frac{8\,y^6}{387420489}+\frac{8\,y^7}{129140163}+\frac{11122\,y^8}{94143178827}+\frac{51818\,y^9}{282429536481}+O(y^{10})~.}
Comparing to the results of \cite{Barrella:2013wja,Chen:2013dxa}, we find agreement through $\Oc(y^8)$.\footnote{We can compare (\ref{MI}) directly to $I_3$ in \cite{Chen:2013dxa}. The mutual information $I_n$, cf. (\ref{IFrelation}), has an overall factor of 1/2 for $n=3$; including the anti-holomorphic part, as they do in \cite{Chen:2013dxa}, contributes an overall factor of 2, so the two factors cancel.} \cite{Chen:2013dxa} only computed through $O(y^8)$, so our term at $O(y^9)$ is new.

At three-loop order, \eqr{f3} implies a nonzero result. Evaluating \eqr{f3} for $p_1=p_2=p(y)$ and $x=x(y)$, we find
\eq{f3ren}{F_{\vac;\,3}(y) = y^{12}\left({13\cdot 2^{10} \over 5^2 \cdot 3^{36} }\right)+O(y^{13}) ~.}
As discussed around \eqr{chenpatt}, the authors of \cite{Chen:2013dxa} computed $F_{\vac;\,3}$ through $O(y^8)$ only, and found zero. We now see that the first contribution appears at $O(y^{12})$. It is remarkable that a computation through $O(y^{11})$ using twist fields would not have revealed the nonzero result!~ This speaks to the different strengths of the twist field method and the sewing expansion that we have performed. 

Finally, nonzero all-loop results at $O(y^{12})$ follow from \eqr{allloop} and \eqr{c44e}:
\es{renall}{F_{\vac ;\,\ell>3} (y)= {y^{12}\over 3^{36}}\cdot\left( {4(3704+590c+125c^2)\over5c(5c+22)}\right)\Bigg|_{c^{1-\ell}}+O(y^{13}) ~.} 

\subsubsection{Other genus-two R\'enyi entropies}\label{v-iib}

Consider the three-interval R\'enyi entropy on the plane with $n=2$. In this case, the replica manifold $\R_{3,2}$ is a genus-two manifold characterized by three moduli that parameterize the positions of the intervals modulo conformal symmetry.\footnote{Besides the case of $n=2$ R\'enyi entropy for two intervals (for which the replica manifold is a torus with complex structure $\t$ given by a known function of the interval length \cite{Lunin:2000yv}), this is the only replica manifold that spans its entire genus $g=(N-1)(n-1)$ moduli space.} Our results for the free energy for $p_1,p_2\ll 1$ and general $x$ can therefore be regarded as (universal contributions to) R\'enyi entropies for the case of three disjoint intervals and $n=2$.  

The period matrix of $\R_{3,2}$ is known in terms of Lauricella functions \cite{Coser:2013qda}. To apply our results, one would first need to understand the relative spacings of intervals that corresponds to $p_1,p_2\ll 1$, by using the map from Schottky space to the period matrix. We do not pursue this geometric picture here. It is clear, however, that not all intervals need to be short, because $x$ is allowed to be general. Thus, we have implicitly provided the first computations of universal contributions to 2D CFT R\'enyi entropies that do not require all intervals to be short. 

One can also consider the case of one interval on the torus with $n=2$. The replica manifold has two moduli, namely, the temperature and interval length. Our methods can again be applied to this case to derive universal contributions to the R\'enyi entropy from the stress tensor sector. This has been done perturbatively in a high or low temperature expansion in \cite{Barrella:2013wja,Chen:2014unl,Datta:2013hba} using different methods that cannot access terms at two-loop and beyond in a large-$c$ expansion, unlike the sewing method here. We note that $p_1$ and $p_2$ as a function of the moduli have been computed perturbatively in \cite{Barrella:2013wja,Chen:2014unl,Datta:2013hba}. We leave the remaining explicit calculation for future work.

\subsection{The separating degeneration limit}\label{v-iii}

An important predecessor of the present work is \cite{Yin:2007gv}, where the relation between $Z_{\rm vac}$ and 3D gravity was first enunciated precisely. Yin tested this relation at genus two, focusing on the separating degeneration limit of the Riemann surface, where $\S$ becomes the union of two tori. Before we probe this region of moduli space with our new results, let us briefly review the work of \cite{Yin:2007gv}. 

For the sake of easy comparison to \cite{Yin:2007gv}, we use his notation in this subsection. We write the elements of the period matrix $\Om$ as
\be
\Om_{11} = \rho~, \quad \Om_{22} = \sigma~, \quad \Om_{12} = \nu~.
\ee
We also define the multiplicative periods
\be
q=e^{2\pi i \rho}~, \quad s= e^{2\pi i \sigma}~, \quad v = 2 \pi i \nu~.
\ee
The separating degeneration limit corresponds to the limit $v\rar 0$ with $(q,s)$ fixed, where $q$ and $s$ parameterize the complex structure of the two tori.  

In the $1/c$ expansion, \cite{Yin:2007gv} computed parts of $F_{\vac;\, 0}, F_{\vac;\, 1}$ and $F_{\vac;\, 2}$ at genus two using a variety of methods, all of which agree:
\vs
\bul Demanding a match to the polar parts of extremal CFT partition functions at low values of $k=c/24$, which are fixed by invariance under the genus-two modular group $Sp(4,\mathbb{Z})$. That this match should hold follows from the definition of extremal CFTs, theories that have no non-trivial Virasoro primaries of dimension less than $k+1$ above the vacuum.

\bul For $F_{\vac;\, 1}$, direct computation of \eqr{grav}.

\bul Direct computation of $Z_{\rm vac}$ written as a sum over bilinears of torus one-point functions of Virasoro vacuum descendants. This is similar to what we do in the present work. 
\vs

Although it is not our focus here, $F_0$ is given by a certain Liouville action whose origins we explain in Appendix \ref{orderc}. In order to write the expressions for $F_{\vac;\;1}$ and $F_{\vac;\,2}$, we must define the holomorphic Eisenstein series, normalized as
\es{eis}{\hat E_n^{\rho} = \sum_{m=1}^{\infty} {m^{n-1} q^m\over 1-q^m}~.}
For $n=2,4$, these hatted versions relate to the usual Eisenstein series as
\es{eis2}{&\hat E^{\rho}_2 = {1-E_2(q)\over 24}\approx q+3q^2+4q^3+O(q^4)\\
&\hat E^{\rho}_4 = {E_4(q)-1\over 240}\approx q+9q^2+28q^3+O(q^4)~.}

The results of \cite{Yin:2007gv}, which we denote $F_{\vac}^{\rm Yin}$, are as follows: in the separating degeneration limit $v\rar 0$, 
\bea\label{Yin1}
F_{\vac;\,1}^{\rm Yin} \!\!\!&=&\!\!\! -\sum_{n=2}^{\infty}\log[(1-q^n)(1-s^n)] + v^2\left({2q\over 1-q}\Eh_2^{\s} + {2s\over 1-s}\Eh_2^{\rho} - 4\Eh_2^{\s}\Eh_2^{\rho}\right)\nonumber\\\!\!\!&+&\!\!\! v^4\Bigg(-\half \left({2q\over 1-q}\Eh_2^{\s} + {2s\over 1-s}\Eh_2^{\rho} - 4\Eh_2^{\s}\Eh_2^{\rho}\right)^2 \\\!\!\!&+&\!\!\! {qs\over 6}\Big(-2(q+s) + 45(q^2+s^2) + 72 qs + 745(q^2s+qs^2) + 3720 q^2s^2\Big)+O(q^4,s^4) \Bigg)\nonumber\\\!\!\!&+&\!\!\!O(v^6)\nonumber
\eea
and\footnote{The semiclassical expansion of the free energy in \cite{Yin:2007gv} was performed in powers of $1/k=24/c$. We expand in $1/c$, and define $F^{\rm Yin}$ according to the $1/c$, rather than the $1/k$, expansion. Thus, our $F_{\vac;\, 2}^{\rm Yin}$ equals $24$ times the $S_2$ found in \cite{Yin:2007gv}.}
\es{Yin2}{
F_{\vac;\,2}^{\rm Yin} &= {2v^2}\left({q\over 1-q}-\Eh_2^{\rho}\right)\left({s\over 1-s}-\Eh_2^{\s}\right)\\&+ 24v^4\left(q^2s^2\left({13\over 36} + {1\over 8}(q+s) - {45\over 16}qs\right)+O(q^4,s^4)\right)+O(v^6)~.}
Note that these are non-perturbative in $q$ and $s$ through $O(v^2)$, and the leading term in $F_1^{\rm Yin}$ is just the sum of one-loop free energies on two tori with periods $q$ and $s$. (Note that in order to recover the $O(v^0)$ piece of $F_{\vac;\,1}^{\rm Yin}$, one relies on \eqr{cx1}.) Expanding everything through $O(q^3s^3)$, we find
\es{Yin3}{F_{\vac;\,1}^{\rm Yin} &= (q^2+q^3+s^2+s^3) \\&- v^2\Big(2 q s (q + s + 3 q s) (2 + 3 q + 3 s + 8 q s)\Big)\\
&+v^4\Bigg({q s\over 6}  \Big(-2 (q+s) + 45 (q^2+s^2)  + 72 q s + 745 (q^2 s  + q s^2) + 3624 q^2 s^2\Big)\Bigg)\\&+O(v^6, q^4, s^4)}
and
\es{Yin4}{
   F_{\vac;\,2}^{\rm Yin} &= 24v^2\left(q^2s^2\left({1\over 3}+ {1\over
2} q+{1\over 2}s+{3\over 4}qs\right) \right) \\
&+ 24v^4\left(q^2s^2\left( {13\over 36} + {1\over 8}(q+s)-{45\over
16}qs  \right)\right) +O(v^6, q^4, s^4)~.}

We are now in a position to extend these results using our computations. As in the previous subsection, our goal is to perform the map $\lbrace p_1,p_2,x \rbrace \mapsto \lbrace q,s,v\rbrace$, express $F_{\vac}$ in these variables, and expand at large $c$. 

In \cite{Yin:2007gv}, the following relations were established:
\es{p1p22}{p_1 &= q\left(1-v^2(2\Eh_2^{\s}) - v^4\left(2(\Eh_2^{\s})^2 + {2\over 3}\Eh_2^{\rho} \Eh_2^{\s}-{1\over 6} \Eh_4^{\s} + {10\over 3}\Eh_2^{\rho} \Eh_4^{\s}\right)+O(v^6)\right)\\
p_2 &= p_1(\sigma\leftrightarrow \rho)~.}
All we need now is to derive $x(q,s,v)$. We do so by inverting one of the Schottky relations in equation \eqr{mltprds}, 
\be\label{gkv}
e^v= x+x\sum_{n,m=1}^{\infty}p_1^np_2^m\sum_{r=-n-m}^{n+m}d(n,m,r)x^r~.
\ee
Note that for $s=q=0$, we have $x=e^v$. So we can write this as $x=e^v + O(q,s,qv, sv)$. The final result for $x$ is rather appealing,
\be\label{xconj}
{x = e^v -4 \Eh_2^{\s} \Eh_2^{\rho}(v^3+v^4)  + O(v^5qs)~.}
\ee
We derive this in Appendix \ref{app-degen}.

Plugging equations \eqr{p1p22} and \eqr{xconj} into \eqr{Z2} enables us to extend the results of \cite{Yin:2007gv} in two ways. First, we can now give the $O(v^4)$ part of the one- and two-loop free energies \eqr{Yin3} and \eqr{Yin4}, respectively, through $O(q^3s^4, q^4s^3)$, not only through $O(q^3s^3)$. Second, and more importantly, we can write down some of the leading terms as $v\rar 0$ for {\it all} loops. 

We find
\bea
F_{\vac;\, 1} \!\!\!&=&\!\!\! F_{\vac;\,1}^{\rm Yin}\nonumber\\\!\!\!&+&\!\!\!v^4\left({1\over 6} q s \left(q^3 (210 + 2764 s + 11865 s^2) + s^3 (210 + 2764 q + 11865 q^2) \right)+O(q^4s^4) \right)\nonumber\\\!\!\!&+&\!\!\!O(v^6)
\eea
and
\be
{F_{\vac;\, 2} =  F_{\vac;\,2}^{\rm Yin}-v^4\Big( q^2s^2\left(14(q^2+s^2) + 283 q s(q+s)\right)+O(q^4s^4) \Big)+O(v^6)~.}
\ee

To read off the free energy at three loops and beyond, we plug \eqr{xconj} into \eqr{f3}--\eqr{c44e}. At three-loop order, we find
\be
{F_{\vac;\, 3} = q^4s^4\left({13312\over 25}v^4 + {86528\over 75} v^6 + O(v^8)\right) + O(q^4s^5v^4, q^5s^4v^4)~.}
\ee
To derive terms at higher orders in $v$ for fixed $q$ and $s$, we would need to expand $x$ beyond $O(v^4)$. Likewise, to derive terms at higher orders in $q$ and $s$ for fixed $v$, we would need to include more terms in the sewing expansion, like $p_1^4p_2^5 C_{4,5}(x)$.
Finally, the all-loop expansion is completed by the terms\footnote{This result should be contrasted with footnote 6 of \cite{Yin:2007gv}.}
\es{allloopyin}{F_{\vac;\, \ell>3} &= q^4s^4\left(v^4+{13\over 6}v^6+O(v^8)\right)\cdot \left({4(3704+590c+125c^2)\over5c(22+5c)}\right)\Big|_{c^{1-\ell}}\\&+ O(q^4s^5v^4, q^5s^4v^4)~.}

\section{Discussion}\label{vi}

We close with a discussion of some open questions, progressing from obvious directions for future work to the more speculative. Some directions for future work were mentioned in the text.
\vs
\bul In the realm of R\'enyi entropy, performing the calculation suggested in section \ref{v-iib} for three intervals would give a satisfying derivation away from a short-interval expansion. In addition, one can straightforwardly apply our results to the case of the $n=2$ R\'enyi entropy for a single interval on the torus, at least in a high or low temperature expansion. One need only perform the map between Schottky coordinates and the temperature and interval length; this map has already been partially performed in \cite{Barrella:2013wja,Chen:2014unl,Datta:2013hba}. No results have yet been derived for the torus case beyond one loop.
\vs
\bul 
One can consider including local operator insertions on $\S$. The sewing procedure remains a sum over sphere correlation functions, now with these extra operator insertions. The operators generate non-vacuum states in the CFT. Taking $\S$ to be a replica manifold, one can thus compute excited-state R\'enyi entropies by the sewing procedure. Such entropies have been computed in CFT using twist-field and holographic methods (e.g. \cite{Alcaraz:2011tn, Astaneh:2013gp, Caputa:2014vaa, Asplund:2014coa, Caputa:2014eta}). As we have tried to demonstrate, the sewing construction is likely to provide a complementary approach that operates at finite $c$, so this seems like an especially worthwhile pursuit. It would be easy, for instance, to read off the $\Oc(c^0)$ terms from the above procedure: these would be predictions for bulk one-loop corrections to the Einstein action evaluated on the ``punctured handlebody''. 
\vs
\bul 
It would be nice to prove that nowhere in the moduli space, except at the separating degeneration point, does the genus-two partition function truncate in a $1/c$ expansion (whereas our method could only access the regime of small $p_1,p_2$). This seems highly likely to be the case. Understanding the structure of the Schottky sum rules in Appendix \ref{app-degen} could also be enlightening.

\vs
\bul 
In our analytic bootstrap of section \ref{iii}, we could equally have used Virasoro conformal blocks rather than global conformal blocks. In this case, the crossed blocks are related to the original blocks by the fusion and braiding matrices. These are known in closed form \cite{Ponsot:2000mt}, so our conclusions can also be phrased in terms of OPE coefficients of Virasoro primaries rather than quasi-primaries. The Virasoro approach is in principle more efficient, as it will fix the four-point function in terms of even fewer pieces of data, and it would be worthwhile to make this precise. An interesting demonstration of this fact comes by way of the $W_3$ correlator $\langle WWWW \rangle$, as computed in \eqr{Lamb}--\eqr{wcoeffs}: up to the norm of $W$, the Virasoro approach would fix $\langle WWWW \rangle$ without having to compute even a single OPE coefficient.
\vs
\bul 
We briefly considered CFTs with higher-spin symmetry; it would be straightforward to extend our computation of $Z_{\rm vac}$ to higher orders for such theories. A more exciting prospect would be to compute the partition function on $\S$ in the presence of insertions that carry higher-spin charge. There is natural motivation for this from holography. In particular, while much work has been done to construct solutions of higher-spin gravity with nonzero higher-spin charge and solid-torus topology \cite{Ammon:2012wc}, there has been no work on building solutions of higher-spin gravity of higher genus and with nonzero higher-spin fields turned on. A subset of such ``higher-spin handlebodies'' would be saddle points of the Euclidean higher-spin gravitational path integral with replica boundary conditions and nonzero higher-spin charge \cite{Perlmutter:2013paa}; accordingly, their action would be expected to match CFT computations of R\'enyi entropy in states with higher-spin charge and/or chemical potentials. This calculation would be analogous to the one peformed in  \cite{Faulkner:2013yia} in the spin-2 case. Constructing such R\'enyi entropies via partition functions on replica manifolds endowed with higher-spin charge, rather than via twist fields \cite{Datta:2014uxa} or Wilson lines \cite{deBoer:2014sna}, would be an interesting application of the replica trick to the higher-spin setting. One might also try to make contact with the ``spin-3 entropy'' of \cite{Hijano:2014sqa}.
\vs
\bul Our results can be used to test the idea that Liouville theory provides an effective description of irrational CFTs with large central charge (see e.g. \cite{Jackson:2014nla} for a recent refinement of this idea and references to earlier work). In particular, the $1/c$ expansion of the genus-two partition function can be checked against a diagrammatic calculation in Liouville theory.

\vs\bul
Upon first glance, the relation between 
\eq{zvac1}{Z_{\vac;\,1} = \sum_{h_1,h_2=0}^{\infty} p_1^{h_1}p_2^{h_2} \lim_{c\rar\infty} C_{h_1,h_2}(x)}
and the bulk graviton determinant \eqr{grav} seems opaque. Nevertheless, these two quantities are equal. Both formulae are written in terms of Schottky data, so it should be possible to find a clean mapping between them. This would be a useful stepping stone to writing down a closed formula for the two-loop contribution to the bulk partition function, in analogy to the determinant \eqr{grav}. In the sewing prescription, the two-loop result simply requires us to sum over the $\Oc(1/c)$ parts of $\Chh$ instead of just the $\Oc(c^0)$ parts. Is there an equally simple prescription in the bulk, and if so do the primitive elements of the Schottky group play a privileged role as they do at one loop?

\vs
\bul 
 Part of the motivation for the present work was the one-loop exactness of the pure-gravity partition function on the solid torus. The current understanding of this result relies on an elegant and simple argument about Virasoro representation theory, which can be understood holographically. It can be derived without recourse to CFT by computing the energies of bulk excitations, or equivalently, by quantizing the phase space given by two copies of $\text{diff}\,S^1/SL(2,\mathbb{R})$ \cite{Maloney:2007ud}. Still, it would be very satisfying to derive this result from a more direct perspective in the bulk. For example, while the solid torus partition function of a pure higher-spin theory is also believed to be one-loop exact, we do not know the analog of diff\,$S^1$ in that context; there should be a more direct argument one can make in the bulk. Understanding this exactness from the perspective of the bulk diagrammatic expansion could provide insights useful for higher genus. 

On the other hand, a perhaps cleverer approach would be to derive the partition function from the $SL(2,\mathbb{R})\times SL(2,\mathbb{R})$ Chern-Simons formulation of 3D gravity. Einstein-Hilbert gravity and Chern-Simons theory are non-perturbatively inequivalent, but it is believed that the semiclassical expansion around a well-defined saddle point can be performed in either formulation. The Chern-Simons approach builds in the topological nature of 3D gravity, whereas the loop expansion of 3D gravity in the metric formulation is no simpler than it is in higher dimensions, despite the absence of propagating bulk degrees of freedom. Presumably, such a computation would be manifestly one-loop exact, in analogy to similar truncations in compact Chern-Simons theory \cite{Jeffrey:1992tk}. (More precisely, all higher-loop effects could be absorbed in a renormalization of the Newton constant.) A Chern-Simons approach would also have the benefit of immediately generalizing to pure higher-spin gravity. The challenge to carrying this out is that both the gauge group and the topology are non-compact. There has been progress in recent years in computing Chern-Simons partition functions for non-compact gauge groups (see e.g. \cite{Dimofte:2009yn}), but the requisite technology does not yet exist for the solid torus. This technology would represent a significant advance in our understanding of 3D gravity.
 
\section*{Acknowledgments}

We thank Matthias Gaberdiel, Jared Kaplan, Christoph Keller, Albion Lawrence, David Poland, David Simmons-Duffin, Herman Verlinde, Roberto Volpato and Xi Yin for helpful discussions. MH, AM, and EP wish to thank the Aspen Center for Physics for hospitality during this work, which was supported in part by National Science Foundation Grant No.\ PHYS-1066293. MH and IGZ were supported in part by the National Science Foundation under CAREER Grant No.\ PHY10-53842. EP was supported in part by funding from the European Research Council under the European Union's Seventh Framework Programme (FP7/2007-2013), ERC Grant agreement STG 279943, ``Strongly Coupled Systems'', and in part by the Department of Energy under Grant No. DE-FG02-91ER40671. IGZ was supported in part by the Department of Energy under Grant No.\ DE-SC0009987.
AM is supported by the National Science and Engineering Research Council of Canada.
 
 \begin{appendix}

\section{An OPE limit of $\Chh$}\label{app-OPE}
The goal in this Appendix is to show that the $x\rar 0$ behavior of $C_{h_1,h-h_1}(x)$ is given, as in \eqr{pc2}, by
\eq{apc2}{\lim_{c\rar\infty}\lim_{x\rar 0}C_{h_1,h-h_1}(x) \sim O(x^{-h})+ {1\over c} O(x^{-h+2}) + \left(\sum_{n=2}^{\infty}{1\over c^n}\right)O(x^{-h+4})} 	
where we again have written $h_2=h-h_1$, and we restrict to $h>h_1>0$. We are ignoring $h_1$- and $h$-dependent coefficients at each order, and displaying only the leading singular behavior at each order in $1/c$. 

The upshot is that \eqr{apc2} follows from considering the $t$-channel OPE limit of the four-point functions that define the $\Chh$. The expansion \eqr{apc2} follows from the $c$-scaling of the OPE coefficients and norms that appear in the conformal block decomposition. The leading $\Oc(c^0)$ term in \eqr{apc2} arises from identity exchange; the leading $\Oc(1/c)$ term, from $T$ exchange; and all higher order terms in $1/c$, from exchange of all other quasi-primaries in the Virasoro identity representation. 

Let us give more detail. Recall that the $C_{h_1,h_2}(x)$ are defined in terms of sums over four-point functions of vertex operators, 
\es{achh}{C_{h_1,h_2}(x) =\sum_{\phi_i,\psi_i\in{{\cal H}_{h_i}}}G_{\phi_1,\psi_1}^{-1}G_{\phi_2,\psi_2}^{-1}\bigg\langle V^{out}(\psi_1,\infty)\;V^{out}(\psi_2,x)\;V^{in}(\phi_2,1)\;V^{in}(\phi_1,0)\bigg\rangle}
where the Hilbert subspaces ${\cal H}_{h_i}$ are spanned by operators of holomorphic dimensions $h_i$. Vertex operators $ V^{out}(\psi,z)$ and $ V^{in}(\phi,z)$ are just chiral CFT operators $\psi(z)$ and $\phi(z)$, respectively, dressed with $z$-dependent factors. In the $z\rar 0$ limit, the dressing factors are finite (cf. section \ref{iv-i}), so we can ignore them. Thus, the $x\rar 0$ limit of $C_{h_1,h_2}(x)$ is simply the $t$-channel limit of a weighted sum over four-point functions of CFT operators, including descendants.

A four-point function of pairwise identical quasi-primary operators of dimensions $h_1$ and $h_2$ can be written in a global conformal block decomposition as
\eq{gblock}{\langle \psi(\infty)\phi(x)\phi(1)\psi(0)\rangle= x^{-h}\sum_{\Oc} {C_{\psi\phi \Oc}^2\over {\cal N}_{\Oc}}x^{h_{\Oc}}{}_2F_1(h_{\Oc},h_{\Oc};2h_{\Oc};x)}
where $C_{\psi\phi \Oc}$ are OPE coefficients, and $h=h_1+h_2$. We are expanding this correlator in the $x\rar 0$ channel. Four-point functions involving secondary operators can be written using derivatives acting on an expression of the above form. For our purposes, we only allow Virasoro descendants of the identity to run in the internal channel: $\Oc\in\lbrace 1,T,\Lambda,\ldots \rbrace$. Indeed, for the purposes of establishing \eqr{apc2}, consideration of the exchange of these three operators alone will be sufficient: that is, we associate the scaling in \eqr{apc2} with specific terms in \eqr{gblock}. Let us write out the first three terms in \eqr{gblock} coming from the Virasoro identity block,  $\Oc\in\lbrace 1,T,\Lambda \rbrace$:
\es{gblock2}{&\langle \psi(\infty)\phi(x)\phi(1)\psi(0)\rangle=\\
 &x^{-h}\left(C_{\psi\phi 1}^2 + {2 \over c}C_{\psi\phi T}^2\,x^2\,{}_2F_1(2,2;4;x) + {10\over c(5c+22)}C_{\psi\phi \Lambda}^2 \,x^4\, {}_2F_1(4,4;8;x)+O(x^6)\right)~.}
We have substituted the explicit operator norms. 

When $\psi$ and $\phi$ are in the same global conformal family, fusion onto the identity is allowed, and $C_{\psi\phi 1}\neq 0$ and independent of $c$. This yields a term of order $\Oc(c^0)$ and $O(x^{-h})$. For a given $(h_1,h_2)$, there is {\it always} such a term in the definition of $C_{h_1,h_2}(x)$, since the latter are defined as a sum over {\it all} correlators involving operators at levels $(h_1,h_2)$. This is most obvious when $h_1=h_2$, because $C_{h_1,h_2}(x)$ will include correlators of four identical operators; but even when $h_1\neq h_2$, the definition of $C_{h_1,h_2}(x)$ includes correlators of arbitrary derivatives of $T$, all of which are in the same global conformal family. For instance, $C_{2,4}(x)$ includes $\langle T(\infty)\, \p^2 T(x) \,\p^2 T(1) \,T(0)\rangle$, which permits fusion onto the identity when $x\rar 0$. There are generically no cancellations among terms in the sum \eqr{achh}. This accounts for the first term on the right-hand side of \eqr{apc2}. This leading behavior was also observed in \cite{Gaberdiel:2010jf}. 

The second term in \eqr{apc2}, at $\Oc(1/c)$, comes from the second term in \eqr{gblock2}. Because $C_{\psi\phi T}$ is $c$-independent\footnote{For instance, when $\psi=\phi$ is a quasi-primary, $C_{\phi\phi T} = h \,{\cal N}_{\phi}$.} and ${\cal N}_T =c/2$, this term contributes at $\Oc(1/c)$ compared to the identity exchange, but not beyond. As explained above, the definition of $C_{h_1,h_2}(x)$ always includes such terms. This accounts for the second term on the right-hand side of \eqr{apc2}. 

The final terms in \eqr{apc2}, at $\Oc(1/c^2)$ and beyond, come from exchange of the level-four quasi-primary $\Lambda$ in \eqr{gblock2}. Because its inverse norm has an infinite expansion in $1/c$, this will contribute a term $O(x^{-h+4})$ to all orders in a $1/c$ expansion, thus accounting for the remaining terms in \eqr{apc2}.

\section{More details on the sewing construction}\label{app-Cs}
\subsection{Operators and norms}\label{app-Cs-i}
In this section, we list the operators and their norms at the first six levels of the Virasoro vacuum representation. To ensure that we have not missed any, it is useful to expand the holomorphic Virasoro vacuum character, $\chi_{\rm vac}$:
\es{chivac}{\chi_{\rm vac} &=\text{Tr}_{\rm vac}(q^{L_0-c/24})\\
&=  q^{-c/24}\prod_{n=2}^{\infty}{1\over 1-q^n}\\
&\approx q^{-c/24}(1+q^2+q^3+2q^4+2q^5+4q^6\ldots)~.}
One can branch $\chi_{\rm vac}$ into global $SL(2,\mathbb{R})$ characters, thereby counting the number of quasi-primary fields. The resulting generating function, call it $\chi_{\rm qp}$, is 
\eq{}{\chi_{\rm qp} = (q^{c/24}\chi_{\rm vac}-1)(1-q) \approx q^2+q^4+2q^6+\ldots.}
In terms of the degeneracy $d(h)$ of all level $h$ operators, the degeneracy of  level $h$ quasi-primaries is $d(h)-d(h-1)$ (for $h>1$). We use the shorthand $\Oc=\Oc(0)$ to denote operators.

\begin{itemize}
\item{Level 2: there is one quasi-primary operator, the stress-energy tensor, $T=L_{-2}|0\rangle$, with norm $\N_T = c/2$.} 
\item{Level 3: there is one secondary operator, $\Oc_3=\p T = L_{-3}|0\rangle$, with norm $\N_{\Oc_3} = 2c$.} 

\item{Level 4: there are two operators,
\be\label{O4}
\cO_4^{(1)}=\L = \left(L_{-2}^2-\frac35L_{-4}\right)|0\rangle~,\quad \cO_4^{(2)} = L_{-4}|0\rangle~.
\ee
The operator $\cO_4^{(1)}$ is the commonly studied quasi-primary often denoted $\L$, and $\cO_4^{(2)}$ is secondary. Their norms are
\be\label{No4}
{\cal N}_{\L}=\frac c2\left(c+\frac{22}5\right),\qquad\quad{\cal N}_{\cO_4^{(2)}}=5c~.
\ee
}
\item{Level 5: there are two operators ,
\be\label{O5}
\cO_5^{(1)} = L_{-1}\left(L_{-2}^2-\frac35L_{-4}\right)|0\rangle~, \quad \cO_5^{(2)} = L_{-5}|0\rangle~,
\ee
where both of them are secondary. Their norms are
\be\label{No5}
{\cal N}_{\cO_5^{(1)}}=4\,c\left(c+\frac{22}5\right),\qquad\quad{\cal N}_{\cO_5^{(2)}}=10c~.
\ee
}
\item{Level 6: there are four operators (all acting on $|0\rangle$),
\bea\label{l6-qps}
\cO_6^{(1)}\!\!\!&=&\!\!\!-\frac{20}{63}L_{-6}-\frac{8}{9}L_{-4}L_{-2}+\frac{5}{9}L_{-3}L_{-3},\\
\cO_6^{(2)}\!\!\!&=&\!\!\!-\frac{(60c+78)}{(70c+29)}L_{-6}-\frac{3(42c+67)}{(70c+29)}L_{-4}L_{-2}+\frac{93}{(70c+29)}L_{-3}L_{-3}+L_{-2}L_{-2}L_{-2},\nonumber\\
\cO_6^{(3)}\!\!\!&=&\!\!\!L_{-1}L_{-1}\big(L_{-2}L_{-2}-\frac35\,L_{-4}\big),\nonumber\\
\cO_6^{(4)}\!\!\!&=&\!\!\!L_{-6},\nonumber
\eea
where $\cO_6^{(1)}$ and $\cO_6^{(2)}$ are quasi-primary, and $\cO_6^{(3)}$ and $\cO_6^{(4)}$ are secondary. Their norms are
\bea\label{No6}
{\cal N}_{\cO_6^{(1)}}\!\!\!&=&\!\!\!\frac 4{63}\,c\left(70c+29\right),\qquad{\cal N}_{\cO_6^{(2)}}=\frac34\,c\,\frac{(2 c-1)\,(5c+22)\,(7c+68)}{(70c+29)},\nonumber\\
{\cal N}_{\cO_6^{(3)}}\!\!\!&=&\!\!\!72\,c\left(c+\frac{22}5\right),\qquad{\cal N}_{\cO_6^{(4)}}=\frac{35}2c~.\nonumber
\eea
}
\end{itemize}
The secondary operators $L_{-n}|0\rangle$, $n>2$ have the form $\partial^{(n-2)}T/(n-2)!$, with norms $n(n^2-1) c/12$. 
\subsection{Four-point functions of vertex operators}\label{app-Cs-ii}
In this section we provide more details on the transformation properties of the vertex operators which were used in computation of the four-point functions $C_{h_1,h_2}(x)$ in subsection \ref{iv-ii}. The final expressions for $C_{h_1,h_2}(x)$ in terms of $x$ are reported in the main text and are not repeated here.

We will need the explicit expressions for the vertex operators at infinity. For $h=2,3$, we have
\es{app-C32}{V^{out}(T,\infty)&=\lim_{z\to\infty}z^4 T(z)\\
V^{out}(\Oc_3,\infty) &= \lim_{z\to\infty} \Big(-z^{6}\partial T(z)-4z^{5}T(z)\Big)~.}
For $h=4$, we have
\es{O41infty}{
V^{out}(\L,\infty)&=\lim_{z\to\infty}z^8 \L(z)\\
V^{out}(\cO_{4}^{(2)},\infty)&=\lim_{z\to\infty}\Big(\frac12z^{8}\partial^2T(z)+5z^{7}\partial T(z)+10z^{6}T(z)\Big)~.}
For $h=5$, we have
\es{O51infty}{
V^{out}(\cO_{5}^{(1)},\infty)&=\lim_{z\to\infty}\Big(-z^{10}-z^9L_1\Big)\,\cO_5^{(1)}(z)\\
V^{out}(\cO_{5}^{(2)},\infty)&=\lim_{z\to\infty}\Big(-\frac16z^{10}\partial^3T(z)-3z^{9}\partial^2T(z)-15z^{8}\partial T(z)-20z^7T(z)\Big)~.
}
Finally, for $h=6$, we have quasi-primary vertex operators
\be
V^{out}(\cO_{6}^{(i)},\infty)=\lim_{z\to\infty}z^{12}\,\cO_6^{(i)}(z),\quad i=\{1,2\}~,
\ee
and secondary vertex operators
\bea\label{O6infty}
V^{out}(\cO_{6}^{(3)},\infty)\!\!\!&=&\!\!\!\lim_{z\to\infty}\Big(z^{12}+z^{11}L_1+\frac{z^{10}}{2}L_1^2\Big)\,\cO_6^{(3)}(z),\nonumber\\
V^{out}(\cO_{6}^{(4)},\infty)\!\!\!&=&\!\!\!\lim_{z\to\infty}\Big(\frac1{24}\,z^{12}\,\partial^4T(z)+\frac76\,z^{11}\,\partial^3T(z)+\,\frac{21}2\,z^{10}\,\partial^2T(z)\nonumber\\&&\quad~ +35\,z^9\,\partial T(z)+35\,z^8T(z)\Big)~.
\eea

With these in hand, we start with $C_{h,2}(x)$. Using the definitions
\es{}{V^{out}(T,x) &= (x-1)^2 \,T(x)\\
V^{in}(T,1) &= (x-1)^2 \,T(1)\\
V^{in}(\Oc,0) &= \Oc(0)}
that follow from section \ref{iv-i}, $C_{h,2}(x)$ takes the general form
\es{}{C_{h,2}(x) = \N_T^{-1}(x-1)^4 \sum_{i=1}^{d(h)}\N_{\Oc_{h}^{(i)}}^{-1} \bigg\langle V^{out}(\cO_{h}^{(i)},\infty)\;T(x)\;T(1)\;\cO_{h}^{(i)}(0)\bigg\rangle~.}

We next consider the functions $C_{h,3}(x)$. For $h=3$ we have
\vspace{-5pt}
\be\label{app-C33}
C_{3,3}(x)=\cl N_{\cO_3}^{-2}\bigg\langle V^{out}(\cO_3,\infty)\;V^{out}(\cO_3,x)\;V^{in}(\cO_3,1)\;V^{in}(\cO_3,0)\bigg\rangle,
\ee
where $V^{out}(\cO_3,\infty)$ is given in \eqr{app-C32} and
\vspace{-5pt}
\bea\label{O3x1}
V^{out}(\cO_3,x)\!\!\!&=&\!\!\!-(x-1)^3\,\partial T(x)-4(x-1)^2\,T(x),\nonumber\\
V^{in}(\cO_3,1)\!\!\!&=&\!\!\!(x-1)^3\partial\,T(1)-4\,(x-1)^2T(1)~.
\eea
The expressions for $C_{4,3}(x)$ and $C_{5,3}(x)$ can then be easily obtained using (\ref{O3x1}) and the vertex operators given above. 

\section{Schottky parameters in the separating degeneration limit}\label{app-degen}

Here, we provide details of the map between Schottky space and the period matrix in the separating degeneration limit considered in section \ref{v-iii}:
\eq{}{\lbrace p_1,p_2,x\rbrace~ \mapsto ~\lbrace q,s,v\rbrace~.}
The final result, perturbative in $v$ but non-perturbative in $q$ and $s$, is
\es{ay}{p_1 &= q\left(1-v^2(2\Eh_2^{\s}) - v^4\left(2(\Eh_2^{\s})^2 + {2\over 3}\Eh_2^{\rho} \Eh_2^{\s}-{1\over 6} \Eh_4^{\s} + {10\over 3}\Eh_2^{\rho} \Eh_4^{\s}\right)+O(v^6)\right)\\
p_2 &= p_1(\sigma\leftrightarrow \rho)\\
x &= e^v -4 \Eh_2^{\s} \Eh_2^{\rho}(v^3+v^4) +  O(v^5qs)~.}
The hatted Eisenstein series were defined in \eqr{eis}. The first two relations were derived in \cite{Yin:2007gv}; here, we will derive the last one. 

The computation is an exercise in series solutions of algebraic equations. We make a series ansatz for $x$,
\eq{x}{x = \sum_{j=0}^{\infty}x_j(q,s) v^j~,}
plug this and the expressions for $p_1$ and $p_2$ in \eqr{ay} into the perturbative Schottky relation
\eq{ayb}{e^v= x+x\sum_{n,m=1}^{\infty}p_1^np_2^m\sum_{r=-n-m}^{n+m}d(n,m,r) x^r~,}
and solve order-by-order for $x_j(q,s)$. 

An immediate question that may occur to the reader is how we are able to obtain a result \eqr{ay} that is non-perturbative in $q$ and $s$, despite only having access to $d(n,m,r)$ to finite order in $(n,m)$. The answer is that we were able to infer various sum rules obeyed by the $d(n,m,r)$ that we believe to hold for all $(n,m)$:
\begin{subequations}\label{idensum}
\bea
&&\sum_{r=-n-m}^{n+m} d(n,m,r) = 0\label{sda}\\
&&\sum_{r=-n-m}^{n+m} d(n,m,r) \,r = 0\label{sdb}\\
&&\sum_{r=-n-m}^{n+m} d(n,m,r) \,r^2 = 0\label{sdc}\\
\sum_{n,m=1}^{\infty}q^ns^m\!\!\!\!\!\!\!\!\!&& \sum_{r=-n-m}^{n+m} d(n,m,r) \,r^3 = 24 \Eh_2^{\rho} \Eh_2^{\s}\label{sdd}\\
&&\sum_{r=-n-m}^{n+m} d(n,m,r)\, r^4= 0\label{sde}
\eea
\end{subequations}
where $\Eh_2^{\rho}$ was defined in \eqr{eis}. We have also found a set of sum rules obeyed by the $c(n,m,|r|)$ that appear in the other two Schottky relations \eqr{mltprds}:
\bea\label{idensumc}
&&\sum_{r=-n-m}^{n+m} c(n,m,|r|) = 0~, \quad  (n,m)\neq (0,0)\nonumber\\ 
&&\sum_{r=-n-m}^{n+m} c(n,m,|r|)\, r^2 = 0 ~, \quad n\neq 0\\
\sum_{m=1}^{\infty} s^m \!\!\!\!\!\!\!\!\!&&\sum_{r=-m}^{m} c(0,m,|r|) \,r^2 = 4 \Eh_2^s~.\nonumber
\eea

We have checked all of these identities through $m=n=7$ using the tables of \cite{Gaberdiel:2010jf}.\footnote{We are grateful to the authors of \cite{Gaberdiel:2010jf} for sharing the relevant Mathematica notebooks.} Actually, we have proven \eqr{idensumc}, as well as \eqr{sda}-\eqr{sdc}. Proof of \eqr{idensumc} follows from comparing a series solution for $p_1$ and $p_2$ using \eqr{mltprds} to the known solution \eqr{ay}, and demanding consistency through $O(v^2)$. Proof of \eqr{sda}-\eqr{sdc} follows from demanding that all three perturbative relations in \eqr{mltprds} yield the same result for $\lbrace p_1,p_2,x\rbrace$: hence, having proven \eqr{idensumc}, we can use these to derive sum rules obeyed by $d(n,m,r)$. Proof of \eqr{sdd} and \eqr{sde} is undoubtedly possible using similar methods.

With these sum rules in hand, we proceed to invert \eqr{ayb}. At $O(v^0)$, we must solve
\eq{v0}{1 = x_0(q,s) + x_0(q,s)\sum_{n,m=1}^{\infty}q^ns^m\sum_{r=-n-m}^{n+m} d(n,m,r) x_0(q,s)^r~.}
But \eqr{sda} implies that $x_0(q,s)=1$ solves \eqref{v0}. At $O(v)$, we must solve
\eq{v1}{1 = x_1(q,s) +x_1(q,s)\sum_{n,m=1}^{\infty}q^ns^m \sum_{r=-n-m}^{n+m}d(n,m,r) r~.}
This time, \eqr{sdb} implies that the second term vanishes, leaving $x_1(q,s)=1$.  The analysis at $O(v^2)$ is nearly identical, and \eqr{sdc} implies $x_2(q,s)=1/2$. 

At $O(v^3)$, the first non-trivial sum appears in the series expansion:
\es{x31}{{1\over 3!} = x_3(q,s) + {1\over 3!}  \sum_{n,m=1}^{\infty}q^ns^m \sum_{r=-n-m}^{n+m}d(n,m,r) r^3~.}
Plugging in \eqr{sdd} leads to
\eq{x3}{x_3(q,s) = {1\over 3!}-4 \Eh_2^{\rho} \Eh_2^{\s}~.}
Finally, at $O(v^4)$, we must solve
\eq{x41}{{1\over 4!} = x_4(q,s) + {1\over 3!}\sum_{n,m=1}^{\infty}q^ns^m\sum_r d(n,m,r) r^3 + {1\over 4!}\sum_{n,m=1}^{\infty}q^ns^m\sum_r d(n,m,r) r^4~.}
The final sum rule \eqr{sde} eliminates the last term, and \eqr{x3} leaves us with
\eq{}{x_4(q,s) = {1\over 4!} - 4 \Eh_2^{\rho} \Eh_2^{\s}~.}
Putting this all together, we find the advertised result in \eqr{ay}. 

We believe that the above sum rules may have interesting applications in other studies of genus-two Riemann surfaces. It would be interesting to understand, for instance, why the sum \eqr{sdd} factorizes. A systematic exploration of all sum rules obeyed by these coefficients would be worthwhile. It seems likely that higher order sum rules may be expressible in terms of holomorphic Eisenstein series \eqr{eis}.


\section{The order-$c$ part of the free energy and the sewing construction}\label{orderc}

In subsections \ref{sewing} and \ref{iv-i}, we reviewed the sewing construction, which expresses the partition function of an arbitrary CFT on a genus-$g$ Riemann surface in terms of $2g$-point functions on the sphere, as illustrated in figure \ref{fig-sewing}. However, the formulas in those subsections, such as \eqref{Zg}, only give the order-$c^0$ and higher (in $1/c$) terms in the free energy $F=-\ln Z$; they miss the order-$c$ term. (This term depends on the full metric on the Riemann surface, not just its complex structure; in other words it depends on the choice of representative of the Weyl class.) This is adequate for the purposes of this paper, since our main interest is in the higher-order terms in $1/c$. However, for completeness, in this appendix we will explain how to obtain the order-$c$ term within the context of the sewing construction.

As an illustrative example, consider an arbitrary CFT on a flat torus with modular parameter $\tau$. The partition function is well-known to be
\begin{equation}\label{Ztau}
Z(\tau) = (p\bar p)^{-c/24}\sum_ip^{h_i}\bar p^{\tilde h_i}=(p\bar p)^{-c/24}\sum_{h,\tilde h}d(h,\tilde h)p^h\bar p^{\tilde h}\,,
\end{equation}
where $p:=e^{2\pi i\tau}$ and $d(h,\tilde h)$ is the multiplicity of operators of weights $h,\tilde h$. To calculate $Z(\tau)$ using \eqref{Zg} (more precisely, its generalization including the antiholomorphic sector), we need to compute the coefficient $C_{h,\tilde h}$. Applying the definitions \eqref{Ch1hg} and \eqref{G-ii}, we find simply $C_{h,\tilde h}=d(h,\tilde h)$. Hence \eqref{Zg} would give $Z(\tau) = \sum_{h,\tilde h}d(h,\tilde h)p^h\bar p^{\tilde h}$; thus, we are missing the factor $(p\bar p)^{-c/24}$.

We proceed with a brief recap of the sewing construction.\footnote{We will closely follow the discussion of the sewing construction in sections 9.3 and 9.4 of \cite{Polchinski1}. However, that reference considered CFTs with vanishing total central charge (in the context of string theory), so the issue we are focusing on here did not arise there. When comparing to that reference, note also that the correlators there are unnormalized, whereas (as throughout this paper) ours are normalized.} For convenience we will assume that the metric on our Riemann surface $M$ is smooth. We now cut it along a circle and glue in two disks, which we call $D_1$ and $D_2$, to obtain a new manifold $M_0$ (which may be connected or disconnected). We choose the metric on these disks in such a way that the metric on the new surface is still smooth. This implies that $D_1$ and $D_2$ can be glued together to make a sphere with a smooth metric, which we'll call $S$. We consider coordinates $z_{1,2}$ on $D_{1,2}$ which, when extended to $S$, obey $z_1=1/z_2$.

The path integral on $M$ can be computed by inserting a complete set of states on the circle where it has been cut. By the state-operator mapping, this is equivalent to inserting a complete set of operators on $D_{1,2}$ at $z_{1,2}=0$, with an appropriate inverse metric $G^{ij}$ (to be determined below) on the space of operators:
\begin{equation}\label{ZM}
Z(M) = Z(M_0)\sum_{i,j}G^{ij}\ev{\mathcal{O}_i^{(z_1)}\mathcal{O}_j^{(z_2)}}_{M_0}\,,
\end{equation}
where the superscripts denote that $\mathcal{O}_i$ is inserted at the origin of the $z_1$ and $\mathcal{O}_j$ at the origin of the $z_2$ coordinate system. More generally, we can start with arbitrary operators $\mathcal{O}_a\cdots$ on $M$:
\begin{equation}\label{opinsertion}
Z(M)\ev{\mathcal{O}_a\cdots}_M=Z(M_0)\sum_{i,j}G^{ij}\ev{\mathcal{O}_a\cdots\mathcal{O}_i^{(z_1)}\mathcal{O}_j^{(z_2)}}_{M_0}\,.
\end{equation}

To fix the inverse metric $G^{ij}$, we consider the case where $M$ happens to include a patch $D_1'$ that is diffeomorphic to $D_1$, with an operator $\mathcal{O}_k$ inserted at $z_1'=0$. Cutting $M$ along the boundary of $D_1'$ and gluing in $D_1$ and $D_2$ yields $M_0=M\cup S$, where the $S$ is covered by coordinates $z_1'$ and $z_2=1/z_1'$. Equation \eqref{opinsertion} then becomes:
\begin{eqnarray}
Z(M)\ev{\mathcal{O}_a\cdots\mathcal{O}_k^{(z_1')}}_M
&=&Z(M_0)\sum_{i,j}G^{ij}\ev{\mathcal{O}_a\cdots\mathcal{O}_k^{(z_1')}\mathcal{O}_i^{(z_1)}\mathcal{O}_j^{(z_2)}}_{M_0} \nonumber\\
&=&Z(M)Z(S)\sum_{i,j}G^{ij}\ev{\mathcal{O}_a\cdots\mathcal{O}_i^{(z_1)}}_M\ev{\mathcal{O}_j^{(z_2)}\mathcal{O}_k^{(z_1')}}_S
\,.
\end{eqnarray}
For this to hold for arbitrary $\mathcal{O}_k$ and arbitrary insertions $\mathcal{O}_a\cdots$, it must be that
\begin{equation}
G_{ij} = Z(S)\ev{\mathcal{O}_i^{(z_1)}\mathcal{O}_j^{(z_2)}}_S = Z(S)\mathcal{G}_{ij}\,,
\end{equation}
where $\mathcal{G}_{ij}$ is the Zamolodchikov metric. 

Now that we have fixed $G^{ij}$, the partition function \eqref{ZM} becomes
\begin{equation}
Z(M) = \frac{Z(M_0)}{Z(S)}\sum_{i,j}\mathcal{G}^{ij}\ev{\mathcal{O}_i^{(z_1)}\mathcal{O}_j^{(z_2)}}_{M_0}\,.
\end{equation}
It is often useful to add a parameter $p$ to the sewing construction, so that the coordinate identification is $z_1z_2=p$. (Even though $p$ can be absorbed in a coordinate transformation on $M_0$, it is useful to fix the coordinate system on $M_0$ and use $p$ to vary the modulus of $M$.) This can be done by replacing $z_2$ by $z_2'$ in the formulas above, and defining $z_2=pz_2'$. We have $\mathcal{O}_j^{(z_2')}=p^{h_j}\bar p^{\tilde h_j}\mathcal{O}_j^{(z_2)}$, so
\begin{equation}\label{ZM2}
Z(M) = \frac{Z(M_0)}{Z(S)}\sum_{i,j}p^{h_j}\bar p^{\tilde h_j}\mathcal{G}^{ij}\ev{\mathcal{O}_i^{(z_1)}\mathcal{O}_j^{(z_2)}}_{M_0}\,.
\end{equation}

Cutting $M$ along $g$ non-contractible cycles, where $g$ is its genus, reduces it to a sphere. This yields the formula \eqref{Zg}, except with a product of sphere partition functions $Z(S_1)\cdots Z(S_g)$ in the denominator. The free energy on any sphere is proportional to $c$, so these factors contribute such a term to $F(M)$. (The coordinate transformation from local coordinates $z_{1,2}$ in the vicinity of each operator insertion to the single coordinate $z$ covering the plane leads to the definition of the operators $V^{out},V^{in}$ explained in subsection \ref{iv-i}.)

To illustrate the application of \eqref{ZM2}, let us return to the example of the flat torus. Set $\beta=\Im\tau$, and let the horizontal cycle have circumference $2\pi$; thus the total area is $4\pi^2\beta$. We will cut it along the horizontal cycle. For $D_1$ and $D_2$ we use unit round hemispheres. Thus $S$ is a round unit sphere, while $M_0$ is a cylinder of circumference $2\pi$ and length $2\pi\beta$ with round endcaps. In the next paragraph we will compute the ratio $Z(M_0)/Z(S)$ using the Liouville action, finding $e^{\pi c\beta/6} = (p\bar p)^{-c/24}$, precisely the prefactor appearing in the expression \eqref{Ztau} for the torus partition function.

In order to compute $Z(M_0)/Z(S)$, we will compute the change in the partition function $Z(M_0)$ under a small change in $\beta$, and then integrate the result up from $\beta=0$ (noting that $M_0|_{\beta=0}=S$). Under a Weyl transformation, $ds^2 = e^{2\omega}d\hat s^2$, the partition function gets transformed by the Liouville action:
\begin{equation}
Z = e^{S_L}\hat Z\,,\qquad
S_L = \frac{c}{24\pi}\int \sqrt{\hat g}\left(\hat g^{ab}\partial_a\omega\partial_b\omega + \hat R\omega\right).
\end{equation}
We will let $d\hat s^2$ be the metric with cylinder length $2\pi\beta$, and $ds^2$ with cylinder length $2\pi(\beta+\delta\beta)$. Hence the Weyl transformation relating them is close to the identity, with $\omega$ of order $\delta\beta$, and we can work to first order in $\omega$. Since the cylinders have the same circumference, $\omega$ vanishes on the cylinder. $\omega$ can also be taken to vanish on, say, the bottom endcap, while on the top endcap it transforms the hemisphere into a hemisphere attached to a thin cylinder of height $2\pi\delta\beta$. On this endcap, $\hat R=2$, so
\begin{equation}
\int\sqrt{\hat g}\hat R\omega = \int\sqrt{\hat g}2\omega = \int\sqrt{g}-\int\sqrt{\hat g} = 4\pi^2\delta\beta\,.
\end{equation}
Hence
\begin{equation}
S_L = \frac{\pi c}{6}\delta\beta\,.
\end{equation}
Integrating from $\beta=0$, we find
\begin{equation}
\ln Z(M_0) = \ln Z(S)+\frac{\pi c}6\beta\,,
\end{equation}
as promised.

\end{appendix}

\bibliographystyle{ssg}
\bibliography{biblio}

\end{document}